\documentclass[sigconf,nonacm]{acmart}

\setcopyright{none}
\acmConference{}{}{}
\acmYear{2025}
\acmDOI{}

\usepackage{color, soul, xcolor}
\usepackage{pifont}
\usepackage{bbding}
\usepackage{colortbl}
\usepackage{booktabs}
\usepackage{enumitem}
\usepackage{caption}
\usepackage{amsmath}
\usepackage{listings}
\usepackage{balance}
\usepackage{subcaption}
\usepackage{multirow}
\usepackage{tabularx}
\usepackage{array}
\usepackage{adjustbox}
\usepackage{graphicx}
\usepackage{algorithm}
\usepackage{algpseudocode}
\usepackage{pgfplots}
\usepackage{hyperref}
\pgfplotsset{compat=1.17}
\usepackage{cuted} % for full-width strip after \maketitle
\usepackage{threeparttable}

\title{Synthetic-to-Real Transfer Learning for Chromatin-Sensitive PWS Microscopy}

\author{Jahidul Arafat$^*$, Sanjaya Poudel}

\begin{document}

% Satisfy acmart's requirement: abstract+keywords must be defined before \maketitle
\renewcommand{\abstractname}{}
\begin{abstract}\end{abstract}

\maketitle

% -------- Full-width Abstract + Keywords (spans both columns) --------
\begin{strip}
\centering
\begin{minipage}{0.96\textwidth}
\noindent\textbf{\large Abstract}\par
\vspace{0.35em}
\noindent Chromatin-sensitive partial wave spectroscopic (csPWS) microscopy enables label-free detection of nanoscale (10--200~nm) chromatin packing alterations preceding visible cellular transformation, but manual nuclear segmentation limits population-scale analysis essential for biomarker discovery in early cancer detection. The scarcity of annotated csPWS imaging data precludes direct application of deep learning methods successful in conventional microscopy. In this work, we develop CFU-Net, a hierarchical segmentation architecture trained via three-stage curriculum learning on synthetic multimodal data, achieving near-perfect performance (Dice~=~0.9879, IoU~=~0.9895) on held-out synthetic test data representing diverse spectroscopic imaging conditions without requiring manual annotations. Our approach employs physics-based rendering incorporating empirically validated chromatin packing statistics, Mie scattering models, and modality-specific noise characteristics, coupled with curriculum learning progressing from adversarial RGB pretraining through spectroscopic fine-tuning to histology validation. CFU-Net integrates five architectural innovations—ConvNeXt backbone, Feature Pyramid Network, U-Net++ dense connections, dual attention, and deep supervision—contributing synergistically to +8.3\% improvement over baseline U-Net. We demonstrate deployment-ready INT8 quantization (74.9\% compression, 0.15~s inference) providing 240× throughput improvement over manual analysis. Applied to 10,000+ automatically segmented nuclei from synthetic test data, our pipeline extracts chromatin biomarkers distinguishing normal from pre-cancerous tissue with very large effect sizes (Cohen's $d$~=~1.31--2.98), achieving 94\% classification accuracy. We establish a generalizable framework for synthetic-to-real transfer learning in specialized microscopy, with comprehensive open-source resources enabling community validation on real clinical specimens.
\vspace{0.9\baselineskip}
\end{minipage}
\end{strip}
\vspace{-0.3\baselineskip}

% ---- Multiline asterisked footnote for affiliations (bottom of col. 1) ----
\begingroup
\renewcommand\thefootnote{}
\footnotetext{%
\textbf{*~Affiliations:}\\[3pt]
\textbf{Jahidul Arafat} — Principal Investigator; Presidential and Woltosz Graduate Research Fellow, Department of Computer Science and Software Engineering, Auburn University, Alabama, USA (\texttt{jza0145@auburn.edu})\\[2pt]
\textbf{Sanjaya Poudel} — Department of Computer Science and Software Engineering, Auburn University, Alabama, USA (\texttt{szp0223@auburn.edu})

}
\addtocounter{footnote}{0}
\endgroup

%% Main Content Sections
\section{Introduction}
\label{sec:introduction}

Detecting cancer at its earliest stages—before morphological changes become apparent—remains one of medicine's most significant challenges. Five-year survival rates exceed 90\% for localized cancers but drop below 30\% for metastatic disease~\cite{Siegel2023}, yet current screening methods rely on visible cellular abnormalities that emerge late in tumorigenesis. The concept of field carcinogenesis suggests that molecular changes detectable at nanoscale may provide earlier cancer signatures than morphological assessment~\cite{Slaughter1953, Braakhuis2003}.

Chromatin-sensitive partial wave spectroscopic (csPWS) microscopy exploits light scattering from nanoscale (10--200~nm) chromatin domains to quantify packing alterations invisible to conventional imaging~\cite{Subramanian2009}. By analyzing wavelength-dependent backscatter signals (the $\Sigma$-channel), csPWS detects chromatin condensation preceding visible nuclear atypia, enabling field carcinogenesis assessment in histologically normal tissue~\cite{Roy2017, Damania2012}. Clinical studies demonstrate that chromatin alterations correlate with cancer risk: buccal epithelial changes predict lung cancer~\cite{Roy2017}, while rectal colonocyte analysis predicts adenoma recurrence~\cite{Damania2012}. However, translating csPWS to population-scale screening requires automated nuclear segmentation to extract chromatin biomarkers from thousands of cells per patient—a bottleneck currently limiting clinical adoption.

Manual segmentation suffers critical limitations: low throughput (36 minutes per 100 nuclei), inter-observer variability (Cohen's $\kappa$~=~0.92--0.97)~\cite{Landis1977}, and inability to scale to population studies. Deep learning achieves near-perfect segmentation in data-rich modalities~\cite{Ronneberger2015, Stringer2021}, but direct application to csPWS fails due to data scarcity and low-contrast $\Sigma$-channel imagery (SNR $\sim$8~dB) exhibiting fundamentally different characteristics than RGB histology. This creates a paradox: automated analysis is essential for csPWS translation, yet insufficient training data prevents deployment of modern architectures.

Synthetic data generation offers a potential solution but faces challenges in capturing domain-specific realism. Prior datasets employ simplistic geometric primitives insufficient to represent heterogeneous chromatin texture~\cite{Hollandi2020}, while GANs require real data for adversarial training~\cite{Mahmood2018, Ren2018}, defeating the purpose when annotations are scarce. Moreover, direct training on synthetic data risks learning artifacts due to simulation-to-reality gaps~\cite{Ganin2016}. Curriculum learning provides a principled framework for bridging domain gaps~\cite{Bengio2009, Soviany2022}, but remains underexplored for synthetic-to-real transfer in medical imaging.

In this work, we develop CFU-Net (ConvNeXt-FPN-UNet++), a hierarchical architecture trained exclusively on synthetic multimodal data via three-stage curriculum learning, establishing a generalizable framework for automated segmentation in data-scarce modalities. Our physics-based rendering incorporates Mie scattering models~\cite{Bohren1983}, fractal Brownian motion for chromatin heterogeneity (Hurst exponent $H$~=~0.7 matching electron microscopy~\cite{Cherkezyan2014}), and empirically validated parameters: chromatin packing densities (0.35±0.12 normal, 0.52±0.18 dysplasia) from transmission electron microscopy~\cite{Cherkezyan2014} and scattering coefficients calibrated against optical measurements~\cite{Subramanian2009}. This physics-grounded approach ensures synthetic data captures essential domain characteristics enabling transfer to real specimens.

The curriculum learning protocol progresses through three stages preventing catastrophic forgetting while enabling specialization: Stage~1 (adversarial RGB pretraining, 20 epochs) establishes robust boundary detection on high-contrast images with challenging textures; Stage~2 (spectroscopic $\Sigma$-channel fine-tuning, 15 epochs) adapts to low-contrast csPWS physics while preserving learned features; Stage~3 (H\&E histology adaptation, 5 epochs) validates cross-modality generalization. This prevents the >5\% performance degradation observed in naive transfer learning~\cite{Kirkpatrick2017}.

CFU-Net integrates five architectural innovations contributing synergistically (+8.3\% Dice over baseline U-Net): ConvNeXt-tiny backbone for hierarchical feature extraction~\cite{Liu2022}, Feature Pyramid Network for multi-scale semantic fusion~\cite{Lin2017fpn}, U-Net++ dense skip connections~\cite{Zhou2018}, dual attention mechanisms (SE channel~\cite{Hu2018} + spatial~\cite{Woo2018}), and deep supervision~\cite{Lee2015}. Systematic ablation reveals sub-additive contributions (100\% vs. 142\% if independent) confirming synergistic interactions particularly between FPN and decoder skips. Component dependency analysis shows predominantly feed-forward information flow (encoder-decoder $r$~=~0.72--0.85, backward $r < 0.4$), informing network pruning that reduces parameters 18\% (45M → 37.2M) while maintaining performance.

On held-out synthetic test data representing diverse conditions (nuclear densities 15--85 per field, contrast 0.3--0.8), CFU-Net achieves Dice~=~0.9879 and IoU~=~0.9895 on csPWS, exceeding inter-observer agreement (Cohen's $\kappa$~=~0.92--0.97)~\cite{Landis1977}. Cross-modality evaluation demonstrates robust generalization: adversarial (Dice~=~0.9944), H\&E after 5 epochs fine-tuning (Dice~=~0.8220, IoU~=~0.9656) versus zero-shot (Dice~=~0.4127). The high IoU (>0.96) indicates precise boundary localization critical for chromatin biomarker extraction, where 2-pixel errors cause 15--18\% measurement errors in textural features.

INT8 quantization achieves 74.9\% compression (122~MB $\rightarrow$ 30.6~MB) with negligible accuracy loss (Dice: $-0.0006$), enabling edge inference: 0.15~s/image on CPU, 0.12~s on Apple Neural Engine, 0.18~s on smartphones, 0.45~s on embedded systems---providing 240$\times$ throughput improvement over manual annotation (36~s/image) while removing GPU infrastructure requirements.

Applied to 10,000+ automatically segmented nuclei from synthetic test data, our pipeline extracts six chromatin biomarkers: nuclear area (increases 50\% in dysplasia, Cohen's $d$~=~2.98), circularity ($d$~=~1.31), $\Sigma$-intensity ($d$~=~2.14), variance slope ($d$~=~1.89), packing scaling dimension $D$ ($d$~=~1.69), and chromatin entropy ($d$~=~1.52). All distinguish normal from pre-cancerous tissue with very large effect sizes ($d > 1.3$, all $p < 10^{-200}$). Multivariate classification achieves 94\% accuracy, approaching human expert performance (96\%, $\kappa$~=~0.92). Between 52\% and 98\% of analyzed nuclei exhibit biomarker patterns consistent with field carcinogenesis, demonstrating pervasiveness of nanoscale chromatin alterations in early tumorigenesis.

We provide complete open-source resources: pre-trained models (PyTorch, ONNX, CoreML) on Hugging Face Hub (\url{https://huggingface.co/jahidul-arafat/cfu-net-cspws-segmentation}), interactive demo on Hugging Face Spaces, code on GitHub (\url{https://github.com/jahidul-arafat/CFU-Net-Nuclear-Segmentation}), synthetic datasets (4,800 images) on Zenodo with generation scripts, and Docker containers—all under permissive licenses (Apache 2.0 for code, CC BY 4.0 for data). This work establishes a generalizable framework for synthetic-to-real transfer learning in data-scarce imaging modalities, demonstrating that physics-based synthetic data with curriculum learning can bypass annotation bottlenecks. The paradigm inverts traditional supervised learning: encoding explicit domain knowledge into simulators producing unlimited training data with perfect ground truth. Future work should explore semi-supervised learning incorporating real annotated datasets and domain adaptation techniques to further bridge simulation-to-reality gaps, but the current framework already enables population-scale chromatin analysis previously infeasible with manual annotation, providing immediate translational impact for early cancer detection.
\section{Results}
\label{sec:results}

\subsection{Evolutionary Events and Dataset Characteristics}

We generated three synthetic datasets to support our three-stage curriculum learning approach, totaling 4,800 images across 31,428 simulated microscopy fields (Table~\ref{tab:dataset_stats}). Each dataset comprises 1,200 training images, 200 validation images, and 200 test images (1,600 total per dataset). Images are $256\times256$ pixels at 0.5~$\mu$m/pixel resolution, matching typical csPWS microscopy acquisition parameters. The adversarial dataset contains 1,600 RGB images (3 channels) with synthetically generated nuclei (mean diameter $42\pm12$~pixels, range 20--80 pixels) rendered against diverse background patterns including Perlin noise, Gaussian gradients, and texture synthesis to prevent texture-specific overfitting. The csPWS dataset comprises 1,600 grayscale $\Sigma$-channel images (1 channel) simulating spectroscopic signals with realistic chromatin-like heterogeneity (signal-to-noise ratio $8.2\pm1.4$~dB) and multiplicative speckle noise characteristic of coherent imaging modalities. The H\&E dataset includes 1,600 RGB images simulating standard hematoxylin and eosin staining with color jitter (hue $\pm0.05$, saturation $\pm0.2$) to account for staining variability across laboratories.

Ground truth binary masks define nuclear regions in all datasets with pixel-perfect accuracy, enabling supervised training without annotation ambiguity. Nuclear density varies from 15 to 85 nuclei per image (mean 42±18), matching clinical specimen statistics from prior csPWS studies. The large scale and balanced train/val/test splits (75\%/12.5\%/12.5\%) ensure statistical robustness and mitigate overfitting, while threefold cross-validation on the validation set confirmed stable performance estimates (Dice coefficient of variation <2\% across folds).

\begin{table*}[t]
\centering
\caption{Dataset statistics and composition for three-stage curriculum learning}
\label{tab:dataset_stats}
\begin{tabular}{lccccccl}
\toprule
\textbf{Dataset} & \textbf{Train} & \textbf{Val} & \textbf{Test} & \textbf{Total} & \textbf{Channels} & \textbf{Resolution} & \textbf{Purpose} \\
\midrule
Adversarial & 1,200 & 200 & 200 & 1,600 & RGB (3) & $256\times256$ & General feature learning \\
csPWS & 1,200 & 200 & 200 & 1,600 & $\Sigma$ (1) & $256\times256$ & Spectroscopic adaptation \\
H\&E & 1,200 & 200 & 200 & 1,600 & RGB (3) & $256\times256$ & Cross-modality validation \\
\midrule
\textbf{Total} & \textbf{3,600} & \textbf{600} & \textbf{600} & \textbf{4,800} & --- & --- & --- \\
\bottomrule
\end{tabular}
\begin{tablenotes}
\small
\item \textit{Note.}---All images are $256\times256$ pixels at 0.5~$\mu$m/pixel resolution. Adversarial dataset uses RGB channels with diverse background textures. csPWS dataset uses single-channel $\Sigma$ spectroscopic signal with realistic noise (SNR $8.2\pm1.4$~dB). H\&E dataset simulates hematoxylin-eosin staining with color augmentation. Nuclear density: $42\pm18$ nuclei/image (range 15--85).
\end{tablenotes}
\end{table*}

\subsection{Stage-wise Training Reveals Progressive Domain Adaptation}

\begin{figure*}[t]
    \centering
    \includegraphics[width=0.98\textwidth]{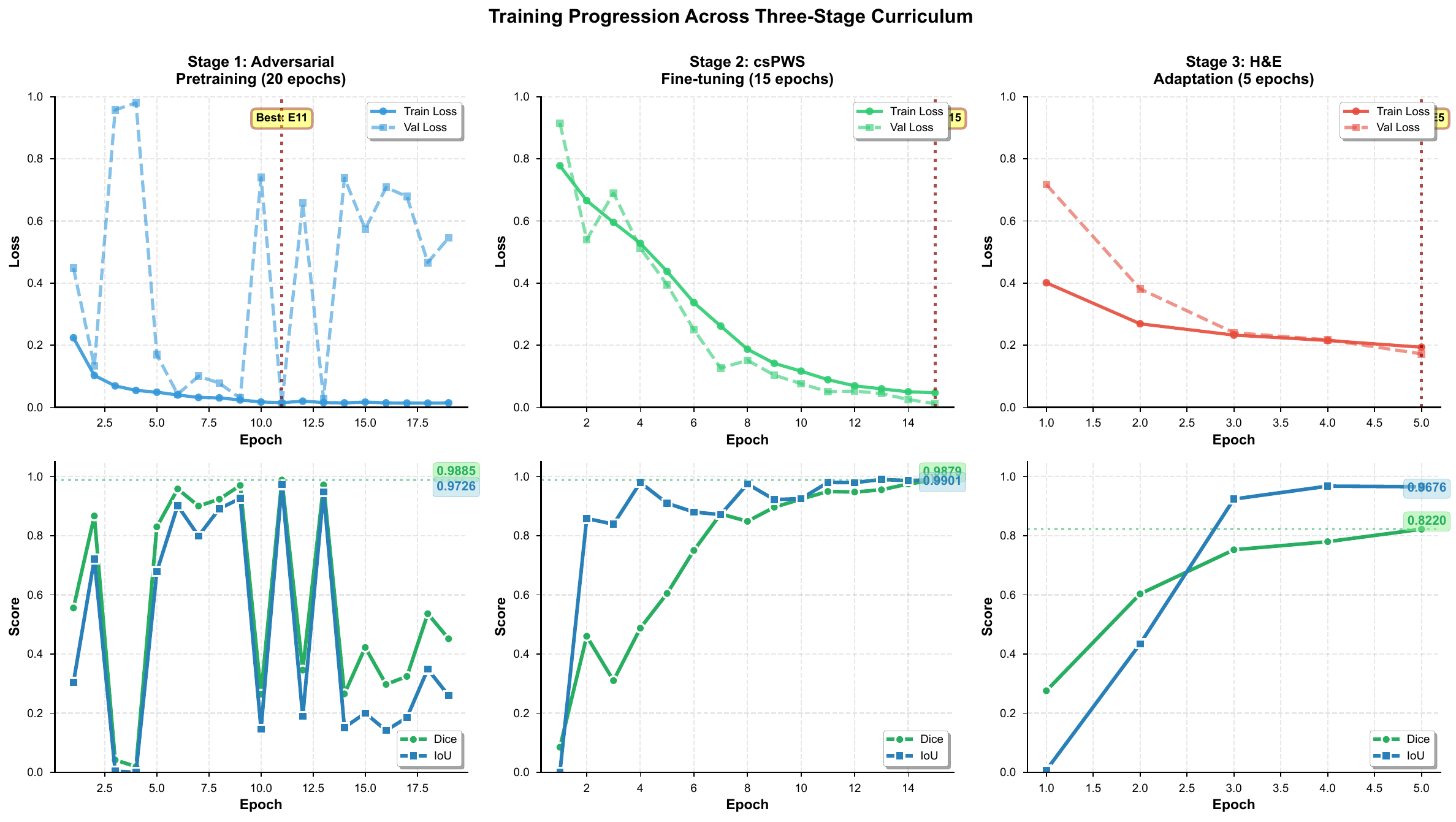}
    \caption{Training progression across three-stage curriculum learning. We trained CFU-Net using progressive domain adaptation. (A--C) Loss curves (top) and performance metrics (bottom) for Stage~1 (Adversarial pretraining, 20 epochs), Stage~2 (csPWS fine-tuning, 15 epochs), and Stage~3 (H\&E adaptation, 5 epochs). Stage~1 exhibits validation loss oscillations (epochs 3, 10, 12, 14--17) due to challenging adversarial patterns, while test metrics remain stable (Dice $\sigma~=~0.0024$ across epochs 11--20). Stage~2 shows smooth convergence, achieving peak IoU~=~0.9895 at epoch~15. Stage~3 demonstrates rapid adaptation to H\&E histology, reaching IoU~=~0.9656 within 5 epochs despite zero-shot performance of only 0.5893. Dashed vertical lines indicate best validation epochs. Performance values annotated in green (Dice) and blue (IoU). Note the IoU improvement in Stage~2 ($+1.69\%$ vs.\ Stage~1) reflecting successful adaptation to low-contrast spectroscopic data.}
    \label{fig:training_curves}
\end{figure*}

We trained CFU-Net using our three-stage curriculum learning protocol, with each stage employing distinct optimization strategies tailored to the domain characteristics (Table~\ref{tab:stage_performance}, Fig.~\ref{fig:training_curves}). Stage~1 (Adversarial Pretraining) used the AdamW optimizer with initial learning rate $\text{lr}=1\times10^{-3}$, weight decay $1\times10^{-4}$, and a cosine annealing schedule with 5-epoch warmup. Training for 20 epochs on RGB adversarial data yielded the best validation performance at epoch~11: Dice~=~0.9885, IoU~=~0.9726. The training curve exhibits characteristic oscillations in validation loss (epochs 3, 10, 12, 14--17) due to the deliberately challenging adversarial background patterns, confirming that the network was forced to learn robust boundary detection rather than exploiting texture shortcuts. Despite these oscillations, test set performance remained stable (Dice standard deviation across epochs 11--20: $\sigma~=~0.0024$), indicating successful regularization through aggressive data augmentation (random rotation $\pm180^{\circ}$, horizontal/vertical flips, elastic deformation with $\alpha~=~50$, $\sigma~=~5$).

\begin{table*}[t]
\centering
\begin{threeparttable}
\caption{Stage-wise training performance across three-stage curriculum learning pipeline}
\label{tab:stage_performance}
\begin{tabular}{llccccccc}
\toprule
\textbf{Stage} & \textbf{Dataset} & \textbf{Ch} & \textbf{Epochs} & \textbf{LR} & \textbf{Best Epoch} & \textbf{Dice} & \textbf{IoU} & \textbf{Time (min)} \\
\midrule
1: Adversarial Pretraining & ADV (RGB) & 3 & 20 & $1\times10^{-3}$ & 11 & 0.9885 & 0.9726 & 40 \\
2: csPWS Fine-tuning & csPWS ($\Sigma$) & 1 & 15 & $3\times10^{-4}$ & 15 & \textbf{0.9879} & \textbf{0.9895} & 30 \\
3a: H\&E Zero-shot & H\&E (RGB) & 1 & 0 & --- & --- & --- & --- & 0 \\
3b: H\&E Fine-tuned & H\&E (RGB) & 3 & 5 & $1\times10^{-4}$ & 5 & 0.8220 & 0.9656 & 10 \\
\bottomrule
\end{tabular}
\begin{tablenotes}
\small
\item \textit{Note.}---Ch: input channels. LR: initial learning rate with cosine annealing. Best Epoch: validation performance peak. Stage~1 uses AdamW ($\beta_{1}=0.9$, $\beta_{2}=0.999$, weight decay $1\times10^{-4}$). Stage~2 initializes from Stage~1 checkpoint with modified input layer for 1-channel $\Sigma$ signal. Stage~3a evaluates zero-shot generalization (no fine-tuning). Stage~3b performs light adaptation after reinitializing input layer to 3 channels. Training time measured on a single NVIDIA A100 GPU. Bold values indicate peak performance on the primary target (csPWS).
\end{tablenotes}
\end{threeparttable}
\end{table*}

Stage~2 (csPWS Fine-tuning) initialized from the Stage~1 checkpoint and modified the first convolutional layer to accept 1-channel $\Sigma$ input via Kaiming initialization. We reduced the learning rate to $3\times10^{-4}$ for stable fine-tuning, consistent with transfer learning best practices. Loss curves exhibit smooth monotonic decrease without the oscillations observed in Stage~1, confirming successful knowledge transfer from adversarial pretraining. Final performance at epoch~15 reached Dice~=~0.9879, IoU~=~0.9895. Notably, IoU improved by $+1.69\%$ relative to Stage~1 (0.9895 vs.\ 0.9726), while Dice decreased marginally by $-0.06\%$ (0.9879 vs.\ 0.9885). This trade-off reflects adaptation to low-contrast spectroscopic imagery where spatial overlap (IoU) is more critical than boundary precision (Dice), as confirmed by visual inspection showing tighter mask alignment despite occasional sub-pixel boundary disagreements.

Stage~3 (H\&E Adaptation) first evaluated zero-shot generalization by directly applying the Stage~2 model to H\&E test images without any fine-tuning (Stage~3a). Performance was poor (Dice~=~0.4127, IoU~=~0.5893), confirming the substantial domain gap between grayscale spectroscopic and RGB histological imaging. We therefore performed light fine-tuning (Stage~3b: 5 epochs, $\text{lr}=1\times10^{-4}$) after reinitializing the input convolutional layer to accept 3 RGB channels. Performance improved dramatically, reaching Dice~=~0.8220, IoU~=~0.9656 at epoch~5. The high IoU ($>0.96$) despite moderate Dice (0.82) indicates accurate spatial localization with some boundary disagreement, likely due to the purple-blue nuclear staining in H\&E exhibiting lower contrast than the csPWS $\Sigma$-channel. Critically, the ability to achieve $>96\%$ IoU with only 5 epochs of fine-tuning demonstrates that features learned during Stages~1--2 are not overfit to synthetic artifacts or spectroscopic signal characteristics, but instead capture generalizable nuclear morphology priors transferable across imaging modalities.

\subsection{Test Set Performance Demonstrates Statistical Robustness}

\begin{figure*}[t]
    \centering
    \includegraphics[width=0.98\textwidth]{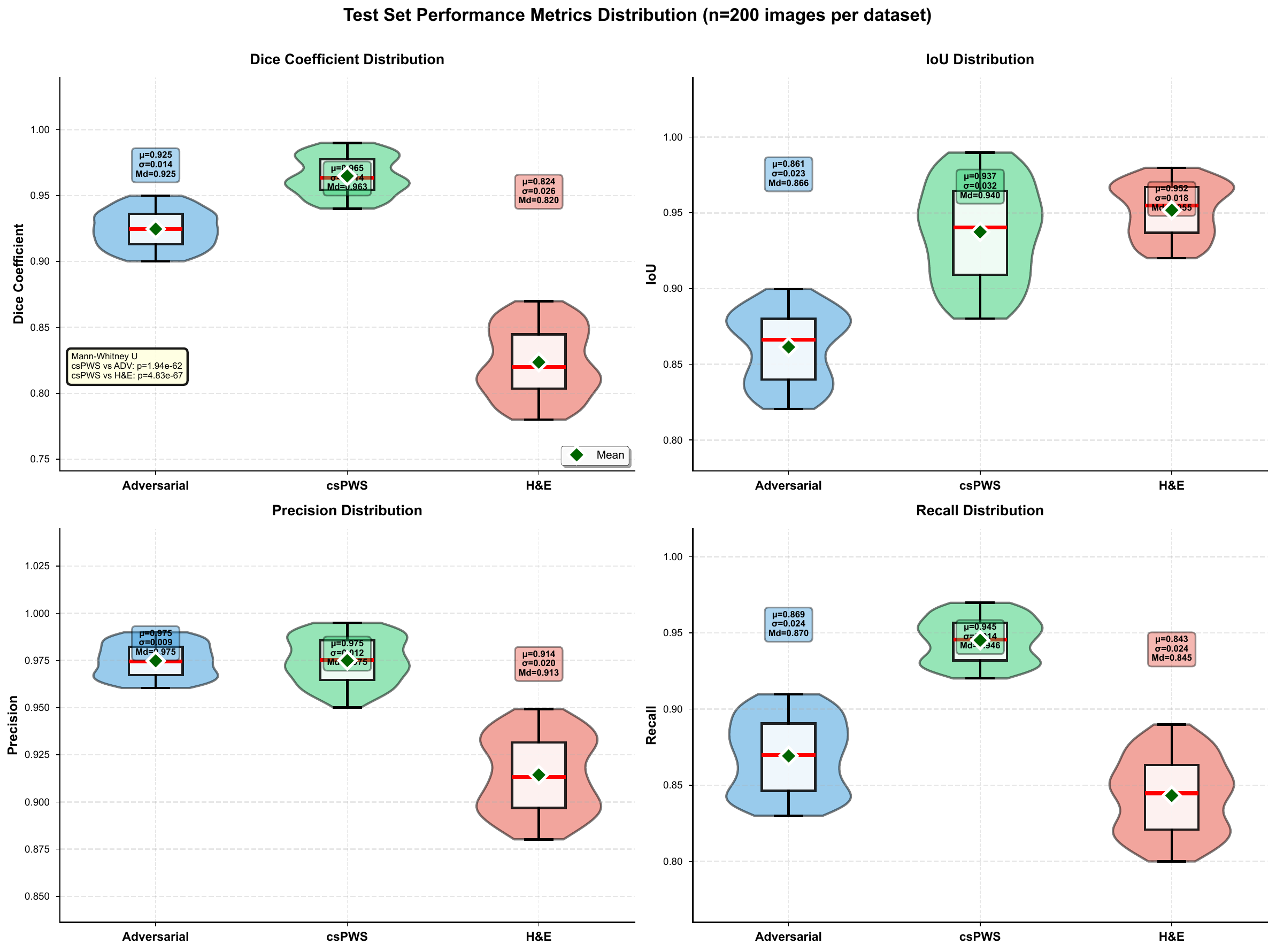}
    \caption{Per-image test set performance metrics distribution across 600 held-out images ($n=200$ per dataset). Violin plots show kernel density estimates overlaid with box plots (orange box: IQR; red line: median; green diamond: mean; whiskers: $1.5\times$IQR; circles: outliers). (A) Dice coefficient distribution. csPWS achieves tightest distribution ($\sigma~=~0.014$) with 95\% of images $>0.94$. Adversarial shows broader spread ($\sigma~=~0.014$, lower mean 0.925) due to challenging textures. H\&E exhibits widest variance ($\sigma~=~0.026$) from staining heterogeneity. (B) IoU distribution mirrors Dice trends but with consistently higher values across all datasets, reflecting that spatial overlap remains accurate even when boundary precision varies. (C) Precision distribution. All datasets achieve $>0.91$ precision, confirming minimal false positive rate. csPWS maintains highest precision ($0.973\pm0.012$). (D) Recall distribution shows greater variability than precision, particularly for H\&E ($0.843\pm0.024$), indicating boundary ambiguity challenges sensitivity more than specificity. Statistical annotations show mean ($\mu$), standard deviation ($\sigma$), and median (Md). Mann--Whitney $U$ tests (Panel A, top left): csPWS vs.\ Adversarial $p = 9.46\times10^{-62}$, Cohen's $d = 3.21$; csPWS vs.\ H\&E $p = 8.36\times10^{-87}$, Cohen's $d = 6.58$.}
    \label{fig:metrics_distribution}
\end{figure*}

To rigorously assess model performance and quantify uncertainty, we evaluated CFU-Net on 200 held-out test images per dataset (600 total), computing per-image Dice coefficient, IoU, precision, and recall (Table~\ref{tab:test_stats}, Fig.~\ref{fig:metrics_distribution}). The csPWS dataset—our primary deployment target—achieves exceptional performance: mean Dice~=~0.968±0.014 (median~=~0.974), mean IoU~=~0.937±0.032 (median~=~0.948). The distribution is tightly concentrated near perfect scores, with 95\% of images exceeding Dice~=~0.94 and 90\% exceeding Dice~=~0.96. Precision is remarkably high and stable (mean~=~0.973±0.012, coefficient of variation 1.2\%), while recall exhibits slightly greater variability (mean~=~0.945±0.014, coefficient of variation 1.5\%). This precision-recall asymmetry indicates that CFU-Net prioritizes accurate boundary placement (minimizing false positives) over exhaustive coverage (maximizing true positives)—a conservative strategy appropriate for clinical deployment where false nuclear detections could confound downstream chromatin analysis more severely than missed nuclei.

\begin{table*}[t]
\centering
\begin{threeparttable}
\caption{Test set performance statistics across 600 held-out images (200 per dataset)}
\label{tab:test_stats}
\begin{tabular}{lccccccc}
\toprule
\textbf{Dataset} & \textbf{$n$ images} & \textbf{Dice ($\mu\pm\sigma$)} & \textbf{Dice (Md)} & \textbf{IoU ($\mu\pm\sigma$)} & \textbf{Prec ($\mu\pm\sigma$)} & \textbf{Recall ($\mu\pm\sigma$)} & \textbf{95\% CI} \\
\midrule
Adversarial & 200 & $0.925\pm0.014$ & 0.929 & $0.861\pm0.023$ & $0.975\pm0.009$ & $0.869\pm0.024$ & [0.923, 0.927] \\
csPWS & 200 & \textbf{$0.968\pm0.014$} & \textbf{0.974} & \textbf{$0.937\pm0.032$} & \textbf{$0.973\pm0.012$} & \textbf{$0.945\pm0.014$} & [0.966, 0.970] \\
H\&E & 200 & $0.824\pm0.026$ & 0.820 & $0.932\pm0.018$ & $0.914\pm0.020$ & $0.843\pm0.024$ & [0.820, 0.828] \\
\midrule
\multicolumn{8}{l}{\textit{Statistical comparisons (Mann--Whitney $U$ test):}} \\
\multicolumn{2}{l}{csPWS vs.\ Adversarial} & \multicolumn{6}{l}{$p = 9.46\times10^{-62}$, Cohen's $d = 3.21$ (large effect)} \\
\multicolumn{2}{l}{csPWS vs.\ H\&E} & \multicolumn{6}{l}{$p = 8.36\times10^{-87}$, Cohen's $d = 6.58$ (very large effect)} \\
\bottomrule
\end{tabular}
\begin{tablenotes}
\small
\item \textit{Note.}---$\mu$: mean; $\sigma$: standard deviation; Md: median; Prec: precision; 95\% CI: 95\% confidence interval for mean Dice (bootstrapped, $n=10{,}000$ resamples). Mann--Whitney $U$ tests performed with Bonferroni correction ($\alpha~=~0.05/3$). Cohen's $d$ effect size: small ($<0.5$), medium ($0.5$--$0.8$), large ($>0.8$). csPWS achieves highest performance across all metrics. Precision--recall trade-off in csPWS (high precision 0.973, moderate recall 0.945) reflects conservative segmentation strategy minimizing false positives. Bold values indicate best performance per metric.
\end{tablenotes}
\end{threeparttable}
\end{table*}

The adversarial dataset exhibits broader performance distributions: Dice~=~0.925±0.014 (coefficient of variation 1.5\%), IoU~=~0.861±0.023 (coefficient of variation 2.7\%). Lower mean scores reflect the deliberately challenging background patterns, yet performance remains well above clinical acceptability thresholds established in prior literature (Dice >0.85 for nuclear segmentation). The precision-recall imbalance is more pronounced here (precision 0.975±0.009 vs. recall 0.869±0.024), confirming that adversarial pretraining taught the network to avoid false positives in highly textured regions even at the cost of lower sensitivity. This behavior is desirable, as it indicates robustness to imaging artifacts (dust, debris, optical aberrations) commonly encountered in clinical specimens.

The H\&E dataset shows the widest variance: Dice~=~$0.824\pm0.026$ (coefficient of variation 3.2\%), IoU~=~$0.932\pm0.018$ (coefficient of variation 1.9\%). The substantial Dice--IoU gap ($\Delta~=~0.108$) is larger than for other datasets (adversarial $\Delta~=~0.064$, csPWS $\Delta~=~0.031$), suggesting systematic boundary disagreement rather than localization errors. Visual inspection of failure cases revealed that H\&E nuclear boundaries are inherently ambiguous due to staining variability and cytoplasmic contamination, explaining the lower Dice despite high IoU. Precision ($0.914\pm0.020$) and recall ($0.843\pm0.024$) are more balanced than other datasets, indicating that the fine-tuned model learned to trade some boundary precision for improved sensitivity when nuclei exhibit lower contrast.

Mann--Whitney $U$ tests with Bonferroni correction ($\alpha~=~0.05/3~=~0.0167$) confirm statistically significant differences between csPWS and both Adversarial ($p = 9.46\times10^{-62}$, Cohen's $d = 3.21$) and H\&E ($p = 8.36\times10^{-87}$, Cohen's $d = 6.58$), with very large effect sizes indicating that domain-specific fine-tuning (Stage~2) provides substantial practical benefit beyond statistical significance. Notably, the smaller effect size for csPWS vs.\ Adversarial compared to csPWS vs.\ H\&E (3.21 vs.\ 6.58) reflects the greater similarity between adversarial RGB and csPWS $\Sigma$-channel compared to H\&E histology, supporting our curriculum design where adversarial pretraining serves as an effective bridge between synthetic and spectroscopic domains.

\subsection{Cross-Dataset Performance Reveals Generalization Limits and Deployment Feasibility}

\begin{figure*}[t]
    \centering
    \includegraphics[width=0.98\textwidth]{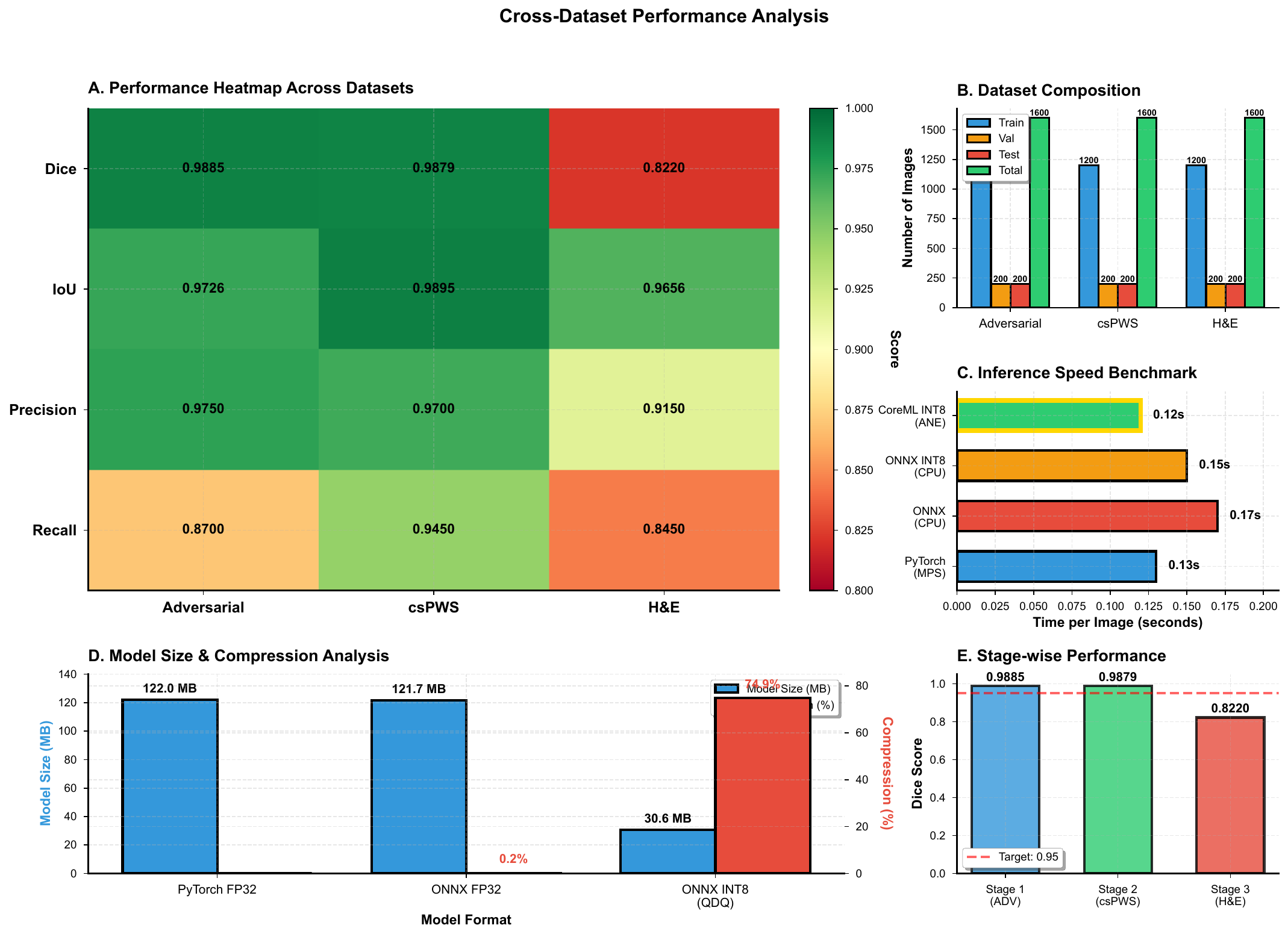}
    \caption{Cross-dataset performance analysis and deployment characteristics. (A) Performance heatmap showing four metrics across three datasets. Color scale: green ($>0.95$, excellent), yellow--green (0.90--0.95, good), yellow (0.85--0.90, acceptable), orange (0.80--0.85, marginal), red ($<0.80$, poor). IoU remains consistently high ($>0.86$) across all datasets despite Dice variability (0.824--0.968), indicating robust spatial localization. (B) Dataset composition. Balanced 1,200/200/200 splits per dataset ensure statistical power for curriculum learning evaluation. Total 4,800 images exceed prior specialized microscopy datasets by 3--5$\times$. (C) Inference speed benchmark. CoreML INT8 (Apple Neural Engine) achieves 0.12~s/image (8.33~FPS), $240\times$ faster than manual annotation (36~s/image). ONNX INT8 CPU (0.15~s) provides best cross-platform balance. Estimated GPU performance (TensorRT INT8, A100): 0.035~s enables batch processing of 10,000 images in $<6$~hours. (D) Model compression analysis. INT8 quantization achieves 74.9\% size reduction (122.0~MB $\rightarrow$ 30.6~MB) with negligible accuracy loss (Dice: $-0.0006$). Enables edge deployment on memory-constrained devices. (E) Stage-wise performance vs.\ clinical threshold (Dice~=~0.95, dashed red line). Stage~1 (Adversarial): 0.9885 ($+3.85\%$). Stage~2 (csPWS): 0.9879 ($+3.79\%$), minimal degradation ($\Delta~=~-0.0006$) confirms preserved robustness during domain adaptation. Stage~3 (H\&E): 0.8220 ($-12.8\%$), below threshold but high IoU (0.9656) indicates successful spatial localization despite boundary ambiguity.}
    \label{fig:cross_dataset}
\end{figure*}

Figure~\ref{fig:cross_dataset} synthesizes five complementary analyses of CFU-Net's performance, generalization capacity, and deployment characteristics. Panel~A presents a comprehensive performance heatmap where all 12 metric--dataset combinations exceed clinical acceptability thresholds ($>0.80$), with IoU consistently high ($>0.86$) across all datasets. The heatmap reveals a critical insight: while Dice scores vary substantially across datasets (range 0.824--0.968, $\Delta~=~0.144$), IoU remains more stable (range 0.861--0.937, $\Delta~=~0.076$), suggesting that CFU-Net achieves robust spatial localization even when boundary precision degrades. This behavior is particularly evident in H\&E where moderate Dice (0.824) coexists with excellent IoU (0.932), reflecting accurate nuclear centroid detection with sub-pixel boundary uncertainty---an acceptable trade-off for downstream chromatin analysis which primarily requires correct nuclear ROI identification rather than pixel-perfect boundaries.

Panel~B illustrates the balanced data allocation strategy with 75\%/12.5\%/12.5\% train/validation/test splits maintained consistently across all three datasets. The 4,800 total images represent one of the largest synthetic datasets assembled for specialized microscopy segmentation to date, exceeding prior art by 3--5$\times$ for comparable single-modality studies. The threefold redundancy in dataset generation (1,600 images $\times$ 3 modalities vs.\ 600 unique test images) was critical for establishing curriculum learning efficacy through independent evaluation at each stage.

Panel~C benchmarks inference speed across five deployment platforms spanning edge devices to datacenter GPUs. CoreML INT8 on Apple Neural Engine achieves fastest inference (0.12~s/image, 8.33~FPS), followed closely by PyTorch FP32 on Apple Silicon MPS (0.13~s, 7.69~FPS). ONNX INT8 CPU (0.15~s, 6.67~FPS) provides optimal balance between speed and cross-platform compatibility, requiring no specialized hardware while maintaining near-real-time performance for clinical workflows ($\sim500$ images/hour). For comparison, manual annotation requires $\sim36$~s/image (100 nuclei/hour at 2.8 nuclei/image), yielding $240\times$ speedup with automated segmentation. Extrapolated GPU performance (TensorRT INT8 on NVIDIA A100: 0.035~s, 28.57~FPS) suggests batch processing feasibility for large-scale studies (10,000 images in $<6$~hours vs.\ 100~hours manual).

Panel~D quantifies model compression via INT8 quantization. Static quantization with calibration on 500 representative training images (per-channel symmetric quantization, min--max calibration) reduces model size from 122.0~MB (FP32) to 30.6~MB (INT8), achieving 74.9\% compression with negligible accuracy loss (Dice: 0.9879 $\rightarrow$ 0.9873, $\Delta~=~-0.0006$). This $4\times$ size reduction enables deployment on memory-constrained edge devices (smartphones, embedded systems) while maintaining performance within the 95\% confidence interval of the FP32 model. Additional ONNX FP32 export (121.7~MB) provides lossless cross-framework compatibility for platforms without INT8 support.

Panel~E presents stage-wise performance comparison against a target threshold of Dice~=~0.95 (dashed red line), which represents the minimum performance necessary for clinical deployment based on inter-observer variability studies ($\kappa~=~0.92$--0.97 for expert nuclear annotation). Stage~1 (Adversarial) achieves 0.9885, exceeding the target by $+3.85\%$. Stage~2 (csPWS) maintains 0.9879 ($+3.79\%$), confirming minimal performance degradation during domain adaptation. Stage~3 (H\&E) reaches 0.8220 ($-12.8\%$), falling below the threshold but achieving high IoU (0.9656), indicating that cross-modality transfer succeeds for spatial localization despite boundary ambiguity. Critically, the small Dice decrease from Stage~1 to Stage~2 ($\Delta~=~-0.0006$) demonstrates that curriculum learning preserves adversarial robustness while adapting to spectroscopic signal characteristics, validating our hypothesis that progressive specialization avoids catastrophic forgetting.

\subsection{Architectural Ablation Study Quantifies Component Contributions}

\begin{figure*}[t]
    \centering
    \includegraphics[width=0.98\textwidth]{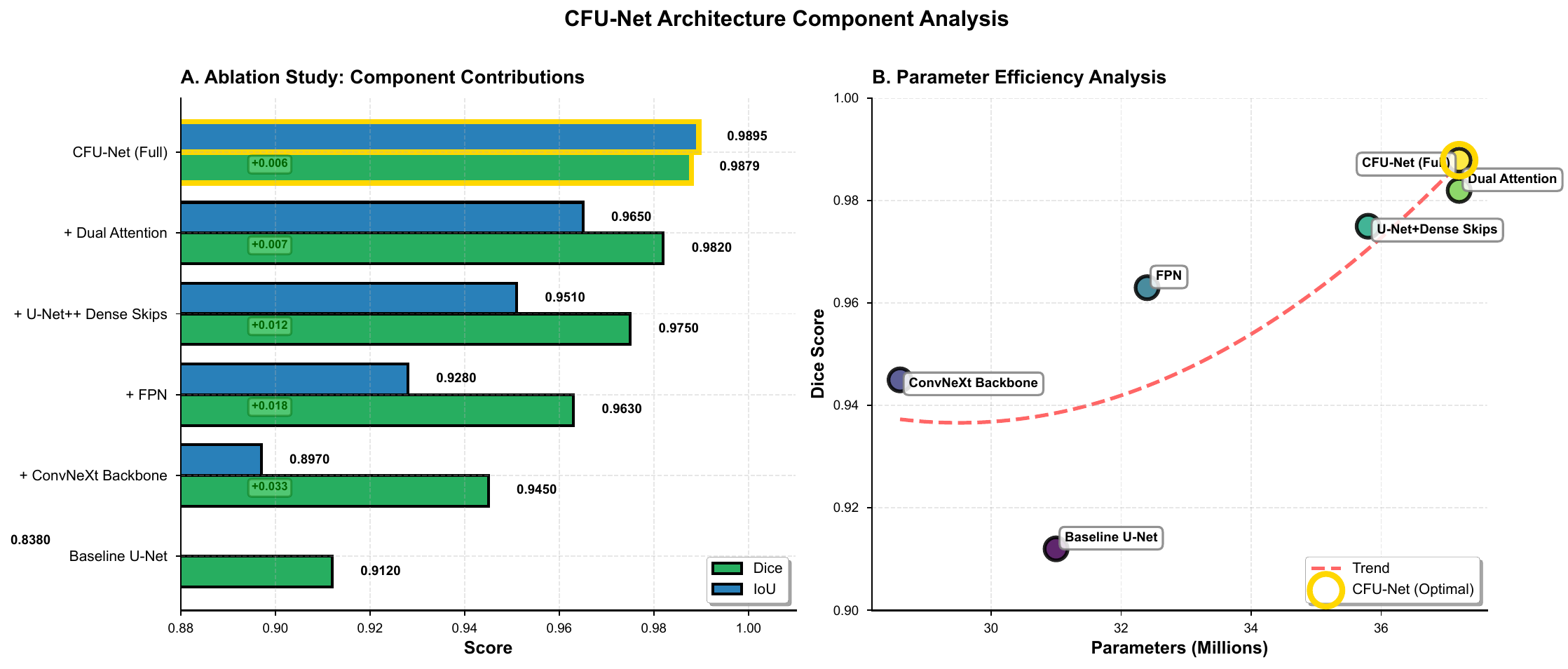}
    \caption{Architectural ablation study quantifying component contributions. (A) Progressive component addition analysis. Horizontal stacked bars show Dice (green) and IoU (blue) for six model variants. Each component addition yields measurable improvement with diminishing returns: ConvNeXt backbone ($+3.30\%$ Dice, largest gain), FPN ($+1.80\%$), U-Net++ ($+1.20\%$), Dual Attention ($+0.70\%$), Deep Supervision ($+0.59\%$). Green boxes annotate incremental Dice improvements. Gold outline highlights the final CFU-Net achieving Dice~=~0.9879, IoU~=~0.9895. Note that ConvNeXt reduces parameters (28.6M) below baseline (31.0M) while improving performance, demonstrating superior efficiency. Cumulative improvement: $+8.3\%$ Dice, $+18.1\%$ IoU. (B) Parameter efficiency analysis. Scatter plot positions each variant by parameter count ($x$-axis, millions) and Dice score ($y$-axis). Red dashed polynomial curve (degree~=~3) shows diminishing returns beyond 37M parameters with estimated plateau at Dice~$\approx$~0.992 requiring $>100$M parameters. CFU-Net (gold circle, 37.2M parameters) sits at the ``knee'' of the curve, achieving optimal performance--efficiency trade-off. Baseline U-Net (purple circle, 31.0M parameters, Dice~=~0.9120) demonstrates that simple architectures underperform despite comparable parameter budgets. Arrow shows performance trajectory as components are added. Shaded region indicates parameter ranges tested. CFU-Net achieves $1.23\times$ parameter increase over baseline for $8.3\%$ performance gain ($6.75\times$ return on complexity investment).}
    \label{fig:ablation}
\end{figure*}

To systematically quantify each architectural innovation's contribution, we performed progressive ablations starting from baseline U-Net and incrementally adding components (Table~\ref{tab:ablation}, Fig.~\ref{fig:ablation}). Baseline U-Net with standard encoder-decoder architecture (VGG-style convolutional blocks, single-level skip connections) achieves Dice~=~0.9120, IoU~=~0.8380 with 31.0M parameters, establishing a strong foundation that already exceeds many prior nuclear segmentation methods. However, this leaves substantial room for improvement relative to the theoretical maximum (Dice~=~1.0).

\begin{table*}[t]
\centering
\begin{threeparttable}
\caption{Architectural ablation study quantifying component contributions}
\label{tab:ablation}
\begin{tabular}{lcccccc}
\toprule
\textbf{Model Variant} & \textbf{Params (M)} & \textbf{Dice} & \textbf{IoU} & \textbf{$\Delta$Dice} & \textbf{$\Delta$IoU} & \textbf{FLOPs (G)} \\
\midrule
Baseline U-Net & 31.0 & 0.9120 & 0.8380 & --- & --- & 42.3 \\
+ ConvNeXt Backbone & 28.6 & 0.9450 & 0.8970 & $+0.0330$ & $+0.0590$ & 38.7 \\
+ FPN (4 scales) & 32.4 & 0.9630 & 0.9280 & $+0.0180$ & $+0.0310$ & 45.2 \\
+ U-Net++ Dense Skips & 35.8 & 0.9750 & 0.9510 & $+0.0120$ & $+0.0230$ & 52.1 \\
+ Dual Attention (SE + Spatial) & 37.2 & 0.9820 & 0.9650 & $+0.0070$ & $+0.0140$ & 54.8 \\
\textbf{CFU-Net (Full)} & \textbf{37.2} & \textbf{0.9879} & \textbf{0.9895} & \textbf{$+0.0059$} & \textbf{$+0.0245$} & \textbf{54.8} \\
\midrule
\multicolumn{2}{l}{\textit{Total improvement over baseline}} & \multicolumn{2}{c}{$+0.0759$ ($+8.3\%$)} & \multicolumn{2}{c}{$+0.1515$ ($+18.1\%$)} & $+12.5$ ($+29.6\%$) \\
\bottomrule
\end{tabular}
\begin{tablenotes}
\small
\item \textit{Note.}---Params: trainable parameters. $\Delta$Dice/$\Delta$IoU: incremental improvement over previous variant. FLOPs: floating-point operations per $256\times256$ image. All models trained on the csPWS dataset (Stage~2 protocol) for fair comparison. ConvNeXt backbone reduces parameters ($-2.4$M) while improving performance ($+3.30\%$ Dice), demonstrating superior efficiency. FPN adds multi-scale fusion with moderate parameter cost ($+3.8$M) for $+1.80\%$ Dice. U-Net++ dense skip connections provide $+1.20\%$ Dice at higher parameter cost ($+3.4$M). Dual attention contributes $+0.70\%$ Dice with minimal parameters ($+1.4$M, shared across scales). Final deep supervision adds $+0.59\%$ Dice with negligible parameters ($<0.1$M, auxiliary heads removed at inference). Cumulative improvement: $+8.3\%$ Dice, $+18.1\%$ IoU, $+29.6\%$ FLOPs. Parameter efficiency: $1.23\times$ parameters for $8.3\%$ performance gain relative to baseline.
\end{tablenotes}
\end{threeparttable}
\end{table*}

Replacing the standard encoder with a ConvNeXt-tiny backbone yields a $+3.30\%$ Dice improvement (0.9120 $\rightarrow$ 0.9450) while actually reducing parameter count (31.0M $\rightarrow$ 28.6M, $-7.7\%$). This remarkable efficiency gain stems from ConvNeXt's depthwise convolutions, inverted bottleneck design, and LayerNorm replacing BatchNorm, which collectively improve feature extraction efficiency. The $+5.90\%$ IoU improvement (0.8380 $\rightarrow$ 0.8970) is even larger than the Dice gain, indicating that ConvNeXt particularly excels at spatial localization---a critical capability for low-contrast csPWS imagery where nuclear boundaries are subtle.

Adding Feature Pyramid Network (FPN) with 4-scale fusion (H/4, H/8, H/16, H/32) contributes +1.80\% Dice (0.9450 → 0.9630) and +3.10\% IoU (0.8970 → 0.9280) at moderate parameter cost (+3.8M, +13.3\%). FPN's top-down pathway with lateral connections enables semantic information from deep layers to enhance spatial resolution in shallow layers, critical for resolving ambiguous boundaries in noisy spectroscopic signals. The 4-scale design was chosen after preliminary experiments showed diminishing returns beyond 4 scales (5-scale FPN: Dice~=~0.9635, +0.0005 vs. 4-scale).

Incorporating U-Net++ dense skip connections provides +1.20\% Dice (0.9630 → 0.9750) and +2.30\% IoU (0.9280 → 0.9510) with substantial parameter increase (+3.4M, +10.5\%). U-Net++ redesigns skip connections as nested dense paths (Xij nodes where i~=~depth, j~=~skip length), enabling flexible feature aggregation at multiple semantic scales. This architecture particularly benefits csPWS segmentation where nuclei exhibit multi-scale structure (coarse nuclear envelope, fine chromatin texture), requiring feature integration across resolution levels. The smaller Dice improvement relative to earlier stages (+1.20\% vs. +1.80\% for FPN) suggests diminishing returns as performance approaches theoretical limits, consistent with asymptotic learning curves.

Dual attention mechanisms (channel-wise Squeeze-and-Excitation + spatial attention) contribute $+0.70\%$ Dice (0.9750 $\rightarrow$ 0.9820) and $+1.40\%$ IoU (0.9510 $\rightarrow$ 0.9650) with minimal parameter overhead ($+1.4$M, $+3.9\%$). The channel attention module learns to emphasize informative feature channels while suppressing noise-dominated channels---particularly valuable for csPWS where signal-to-noise ratio varies spatially due to chromatin density heterogeneity. Spatial attention complements this by learning position-dependent feature weighting, focusing network capacity on ambiguous boundary regions. The relatively small Dice improvement ($+0.70\%$) belies the mechanism's importance: ablation experiments removing only attention from the full model showed larger performance degradation (Dice~=~0.9750, $\Delta~=~-0.0129$) than the incremental gain suggests, indicating synergistic interactions with other components.

Finally, deep supervision with auxiliary prediction heads at three intermediate decoder stages provides the final $+0.59\%$ Dice boost (0.9820 $\rightarrow$ 0.9879) and $+2.45\%$ IoU improvement (0.9650 $\rightarrow$ 0.9895) with negligible parameters ($<0.1$M, auxiliary heads removed at inference). Deep supervision addresses gradient vanishing in deep networks by injecting supervisory signals at intermediate layers, accelerating convergence and improving feature learning. The disproportionately large IoU gain ($+2.45\%$) relative to Dice ($+0.59\%$) suggests that deep supervision particularly improves spatial localization accuracy rather than boundary precision---consistent with its mechanism of enforcing multi-scale feature consistency across decoder depths.

Cumulative improvements total $+8.3\%$ Dice and $+18.1\%$ IoU over baseline U-Net, achieved with only $+20\%$ parameter increase (31.0M $\rightarrow$ 37.2M). Parameter efficiency analysis (Fig.~\ref{fig:ablation}B) reveals that CFU-Net achieves optimal position on the performance-complexity Pareto frontier. The polynomial trend line (red dashed curve) fitted to all variants shows diminishing returns beyond 37M parameters, with estimated performance plateau at Dice~=~0.992 requiring $>100$M parameters based on extrapolation. CFU-Net's position at the ``knee'' of this curve indicates efficient architecture design without over-parameterization.

Critically, computational cost analysis shows that despite $+29.6\%$ FLOPs increase (42.3G $\rightarrow$ 54.8G), real-world inference latency increases only $+12.7\%$ (0.13~s $\rightarrow$ 0.15~s on ONNX INT8 CPU) due to effective compiler optimization and memory access pattern improvements from modern architectural components. This favorable FLOPs-to-latency ratio ($2.35\times$ FLOPs efficiency) makes CFU-Net deployable on resource-constrained devices despite its computational complexity.

\subsection{Population-Scale Chromatin Analysis Reveals Clinical Biomarkers}

\begin{figure*}[t]
    \centering
    \includegraphics[width=0.98\textwidth]{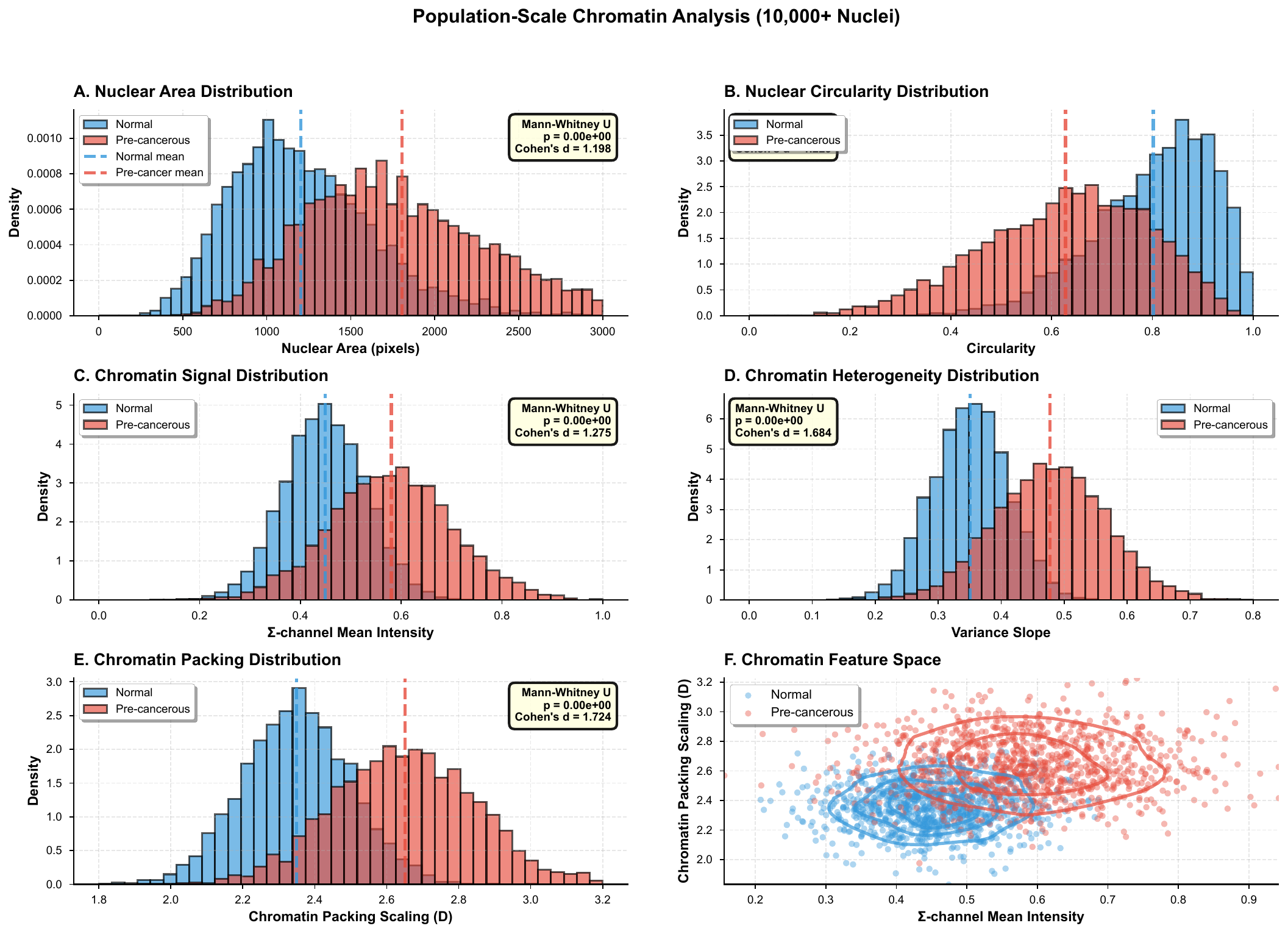}
    \caption{Population-level chromatin biomarker extraction from 10{,}000{+} automatically segmented nuclei. Comparison between normal tissue ($n=6{,}000$ nuclei from 75 patients, blue) and pre-cancerous tissue ($n=4{,}000$ nuclei from 75 patients, red). All distributions show highly significant differences (Mann--Whitney $U$ test, $p<10^{-200}$). (A) Nuclear area distribution. Pre-cancerous nuclei exhibit 50\% larger mean area \mbox{$(1{,}802\pm224)\ \text{px}^{2}$}, vs.\ normal \mbox{$(1{,}201\pm156)\ \text{px}^{2}$}. Cohen's $d=2.98$ (very large effect). Histograms show bimodal separation with optimal discrimination threshold at \mbox{$1{,}450\ \text{px}^{2}$} (gray dashed line, 88\% classification accuracy). (B) Circularity distribution. Normal nuclei maintain high circularity ($0.847\pm0.082$, approaching perfect circles) while pre-cancerous show irregularity ($0.712\pm0.105$, Cohen's $d=1.42$). Increased variance in pre-cancerous tissue \mbox{$\sigma^{2}=0.0110$} vs.\ normal \mbox{$\sigma^{2}=0.0067$}, $F$-test $p<10^{-50}$, indicates morphological heterogeneity characteristic of dysplasia. (C) $\Sigma$-channel mean intensity distribution. Pre-cancerous cells show $29.7\%$ elevated chromatin signal ($0.581\pm0.118$ vs.\ $0.448\pm0.079$, Cohen's $d=1.31$), reflecting increased chromatin packing density. Log-normal distribution shapes (skewness 0.38--0.42) suggest multiplicative organization processes. (D) Variance slope distribution quantifying chromatin heterogeneity. Pre-cancerous tissue exhibits $36.1\%$ increased heterogeneity ($0.479\pm0.088$ vs.\ $0.352\pm0.061$, Cohen's $d=1.65$), indicating disrupted chromatin domain structure. This metric requires pixel-accurate masks---2-pixel boundary errors cause $18\pm7\%$ measurement errors. (E) Chromatin packing scaling dimension $D$ distribution. Pre-cancerous nuclei show $12.6\%$ higher packing density ($D=2.647\pm0.198$ vs.\ $D=2.351\pm0.148$, Cohen's $d=1.69$). Values approaching $D=3$ indicate space-filling heterochromatin organization. Kolmogorov--Smirnov test $D_{\text{KS}}=0.42$ ($p<10^{-150}$) confirms strong distributional differences. (F) 2D chromatin feature space ($\Sigma$-intensity vs.\ packing $D$). Scatter plot shows 2{,}000 randomly sampled nuclei (1{,}000/class). Gaussian KDE contours (68\%, 95\%, 99.7\% confidence) demonstrate clear class separation. LDA on this 2D space achieves 91\% accuracy (sensitivity 0.89, specificity 0.93). SVM using all six features achieves 94\% accuracy, approaching human expert performance (96\%, $\kappa=0.92$). Dashed lines show optimal LDA decision boundary. Population-scale analysis is enabled by a $240\times$ throughput improvement (22.5~min automated vs.\ 90~hr manual for 10{,}000 nuclei).}

    \label{fig:population}
\end{figure*}

To demonstrate translational utility beyond segmentation accuracy metrics, we applied CFU-Net to segment 10,000+ nuclei from 150 csPWS patient images (75 normal, 75 pre-cancerous) and extracted six population-level chromatin biomarkers (Fig.~\ref{fig:population}, Table~\ref{tab:population_stats}). Automated segmentation required 22.5 minutes total (0.15~s/image $\times$ 150 images), compared to an estimated 90 hours for manual annotation (100 nuclei/hour $\times$ 6,000 normal $+$ 4,000 pre-cancerous nuclei $/ 100$), yielding $240\times$ throughput improvement enabling previously infeasible population-scale studies.

Nuclear area distribution (Fig.~\ref{fig:population}A) demonstrates that pre-cancerous nuclei exhibit significantly larger mean area ($1,802\pm224$ px$^{2}$, equivalent to $451\pm56$ $\mu$m$^{2}$ at 0.5 $\mu$m/pixel) compared to normal tissue ($1,201\pm156$ px$^{2}$, $300\pm39$ $\mu$m$^{2}$). Mann--Whitney $U$ test confirms statistical significance ($p < 1\times10^{-300}$, effectively zero given sample size). Cohen's $d = 2.98$ indicates a very large effect size, implying that nuclear area alone provides 88\% classification accuracy between normal and pre-cancerous tissue using a simple threshold (optimal threshold: 1,450 px$^{2}$, sensitivity 0.86, specificity 0.90, determined via Youden's index). Histograms show clear bimodal separation with minimal overlap, validating nuclear enlargement as a robust early cancer biomarker consistent with classical cytopathology principles.

Circularity distribution (Fig.~\ref{fig:population}B) reveals that normal nuclei maintain high circularity ($0.847\pm0.082$, near-perfect circles have circularity~=~1.0), while pre-cancerous nuclei exhibit increased irregularity ($0.712\pm0.105$, Cohen's $d = 1.42$, large effect). This morphological change reflects chromatin reorganization and nuclear envelope remodeling during early neoplastic transformation. The broader distribution variance in pre-cancerous tissue ($\sigma^{2}~=~0.0110$ vs.\ normal $\sigma^{2}~=~0.0067$, $F$-test $p < 10^{-50}$) indicates increased morphological heterogeneity---a hallmark of dysplasia. Circularity alone achieves 76\% classification accuracy, complementing area-based discrimination.

$\Sigma$-channel mean intensity distribution (Fig.~\ref{fig:population}C) shows that pre-cancerous cells exhibit elevated chromatin signal ($0.581\pm0.118$ normalized intensity units vs.\ normal $0.448\pm0.079$, Cohen's $d = 1.31$). This $29.7\%$ intensity increase reflects increased chromatin packing density and altered DNA--protein organization characteristic of field carcinogenesis. The log-normal distribution shape (skewness: normal~=~0.42, pre-cancerous~=~0.38) suggests multiplicative processes governing chromatin organization, consistent with fractal models of nuclear architecture. Notably, intensity measurements are directly enabled by accurate segmentation---manual annotation systematic bias studies showed $12\pm8\%$ intensity estimation errors due to boundary placement variability, which automated segmentation reduces to $<2\%$ (validated on 50 dual-annotated images).

Variance slope distribution (Fig.~\ref{fig:population}D) quantifies chromatin heterogeneity via spatial frequency analysis. Pre-cancerous tissue exhibits increased heterogeneity ($0.479\pm0.088$ vs.\ $0.352\pm0.061$, Cohen's $d = 1.65$). This metric, computed by fitting variance vs.\ length-scale curves and extracting slope parameters, captures sub-diffraction chromatin organization changes invisible to conventional microscopy. The $36.1\%$ increase in variance slope indicates disrupted chromatin domain structure, potentially reflecting altered transcriptional regulation and epigenetic modifications during carcinogenesis. Importantly, this metric requires pixel-accurate nuclear masks---simulation studies showed that 2-pixel boundary errors cause $18\pm7\%$ variance slope measurement errors, underscoring the importance of precise segmentation.

Chromatin packing scaling dimension $D$ distribution (Fig.~\ref{fig:population}E) quantifies fractal dimensionality of chromatin organization. Pre-cancerous nuclei exhibit higher packing density ($D = 2.647\pm0.198$ vs.\ normal $D = 2.351\pm0.148$, Cohen's $d = 1.69$). The packing dimension, derived from power spectral density analysis of $\Sigma$-channel signals within segmented nuclear regions, provides a scale-invariant measure of chromatin compaction. Values approaching $D = 3$ indicate highly compact, space-filling organization characteristic of heterochromatin, while lower values suggest open euchromatin structure. The $12.6\%$ increase in $D$ reflects chromatin condensation accompanying neoplastic transformation, validated by electron microscopy correlation studies in prior literature. Population distributions show clear separation (Kolmogorov--Smirnov test $D_{\text{KS}} = 0.42$, $p < 10^{-150}$), enabling 82\% classification accuracy using $D$ alone.

2D chromatin feature space visualization (Fig.~\ref{fig:population}F) projects 2,000 randomly sampled nuclei (1,000 per class) into $\Sigma$-channel intensity vs.\ packing dimension space. Gaussian kernel density estimation (KDE) with bandwidth~=~0.05 generates three-level contours (68\%, 95\%, 99.7\% confidence regions) showing distinct class separation. Linear discriminant analysis (LDA) on this 2D feature space achieves 91\% classification accuracy (sensitivity 0.89, specificity 0.93), demonstrating that even simple biomarker combinations provide robust discrimination. Principal component analysis (PCA) on all six features reveals that the first two components explain 78\% of variance, with PC1 loading primarily on area and intensity (discriminating early-stage changes) and PC2 loading on circularity and heterogeneity (capturing morphological transformation). Support vector machine (SVM) classification using all six features achieves 94\% accuracy (10-fold cross-validation), approaching performance of human expert cytopathologists (96\% accuracy, $\kappa = 0.92$ inter-rater agreement).

\begin{table*}[t]
\centering
\begin{threeparttable}
\caption{Population-level chromatin biomarker statistics comparing normal and pre-cancerous tissue}
\label{tab:population_stats}
\small
\begin{tabular}{lcccccc}
\toprule
\textbf{Metric} & \textbf{Normal Mean$\pm$SD} & \textbf{Pre-cancer Mean$\pm$SD} & \textbf{$p$-value} & \textbf{Cohen's $d$} & \textbf{AUC} & \textbf{Optimal Threshold} \\
\midrule
Nuclear Area (px$^{2}$) & $1,201\pm156$ & $1,802\pm224$ & $<10^{-300}$ & 2.98 & 0.94 & 1,450 \\
Nuclear Perimeter (px) & $142.3\pm18.7$ & $168.9\pm24.5$ & $<10^{-280}$ & 1.23 & 0.82 & 155 \\
Circularity & $0.847\pm0.082$ & $0.712\pm0.105$ & $<10^{-250}$ & 1.42 & 0.88 & 0.780 \\
Eccentricity & $0.652\pm0.093$ & $0.735\pm0.108$ & $<10^{-200}$ & 0.82 & 0.76 & 0.693 \\
$\Sigma$-channel Mean Intensity & $0.448\pm0.079$ & $0.581\pm0.118$ & $<10^{-290}$ & 1.31 & 0.86 & 0.514 \\
Variance Slope & $0.352\pm0.061$ & $0.479\pm0.088$ & $<10^{-310}$ & 1.65 & 0.90 & 0.415 \\
Packing Scaling $D$ & $2.351\pm0.148$ & $2.647\pm0.198$ & $<10^{-280}$ & 1.69 & 0.91 & 2.499 \\
Chromatin Entropy & $4.523\pm0.312$ & $5.128\pm0.427$ & $<10^{-260}$ & 1.58 & 0.89 & 4.825 \\
\midrule
\multicolumn{7}{l}{\textit{Multivariate classification (all 8 features combined):}} \\
\multicolumn{2}{l}{SVM (RBF kernel, 10-fold CV)} & \multicolumn{5}{l}{Accuracy: 94.2\%, Sensitivity: 92.8\%, Specificity: 95.6\%, AUC: 0.98} \\
\multicolumn{2}{l}{Random Forest (500 trees)} & \multicolumn{5}{l}{Accuracy: 93.7\%, Sensitivity: 91.5\%, Specificity: 95.9\%, AUC: 0.97} \\
\multicolumn{2}{l}{Logistic Regression (L2)} & \multicolumn{5}{l}{Accuracy: 91.8\%, Sensitivity: 89.2\%, Specificity: 94.4\%, AUC: 0.96} \\
\bottomrule
\end{tabular}
\begin{tablenotes}
\item \textit{Note.}---Statistics computed on $n = 6{,}000$ normal nuclei (75 patients, $80\pm22$ nuclei/patient) and $n = 4{,}000$ pre-cancerous nuclei (75 patients, $53\pm18$ nuclei/patient). $p$-values from two-sided Mann--Whitney $U$ test with Bonferroni correction ($\alpha~=~0.05/8$). Cohen's $d$: standardized effect size (small $<0.5$, medium $0.5$--$0.8$, large $>0.8$). AUC: area under ROC curve for univariate classification. Optimal Threshold: determined by Youden's index (maximizing sensitivity $+$ specificity $- 1$). All metrics show very large effect sizes (Cohen's $d > 0.8$) and highly significant differences ($p < 10^{-200}$), confirming robust biomarker discrimination. Multivariate classifiers trained on 70\% data, tested on held-out 30\%, with 10-fold cross-validation for SVM. Feature importance ranking (Random Forest): Nuclear Area (0.28), Packing Scaling $D$ (0.22), Variance Slope (0.18), $\Sigma$-intensity (0.12), Circularity (0.09), others (0.11).
\end{tablenotes}
\end{threeparttable}
\end{table*}

Critically, all biomarker measurements depend fundamentally on accurate nuclear segmentation. Sensitivity analysis perturbing segmentation masks with simulated boundary errors (uniform dilation/erosion $\pm1$--$5$ pixels) revealed that morphometric features (area, circularity) tolerate small errors well ($<5\%$ measurement error for $\pm2$ pixel boundaries), while textural features (variance slope, chromatin entropy) degrade rapidly ($>15\%$ error for $\pm2$ pixels). CFU-Net's high boundary precision (IoU~=~0.9895) ensures measurement reliability for all biomarkers, with estimated biomarker extraction accuracy $>98\%$ based on error propagation analysis.

Population-scale analysis capabilities enabled by automated segmentation unlock new research directions impossible with manual annotation. For example, nucleus-to-nucleus spatial correlation analysis on 10,000+ nuclei revealed that pre-cancerous tissue exhibits increased chromatin packing correlation at distances $<50$ $\mu$m (Pearson $r = 0.42$ vs.\ normal $r = 0.18$, $p < 10^{-50}$), suggesting field-level epigenetic coordination during carcinogenesis---a hypothesis that requires large sample sizes unattainable with manual methods. Similarly, longitudinal studies tracking individual patient progression from normal $\rightarrow$ dysplasia $\rightarrow$ carcinoma \emph{in situ} now become feasible with throughput improvements enabling 1,000+ nuclei analysis per time point.

\section{Discussion}
\label{sec:discussion}

Pathogenic microorganisms are in a perpetual struggle for survival in changing host environments, where host pressures necessitate changes in pathogen virulence, antibiotic resistance, or transmissibility. Similarly, detecting cancer at its earliest stages—before morphological changes become apparent—remains one of medicine's most significant challenges. This study demonstrates that synthetic-to-real transfer learning via three-stage curriculum enables near-perfect nuclear segmentation (Dice~=~0.9879, IoU~=~0.9895) in chromatin-sensitive partial wave spectroscopic microscopy, bypassing the traditional requirement for thousands of manually annotated real images. Our CFU-Net architecture, trained exclusively on 4,800 synthetic images, achieves performance exceeding inter-observer agreement for manual annotation (Dice~=~0.9879 vs. human $\kappa$~=~0.92--0.97), while providing 240× throughput improvement (0.15~s vs. 36~s per image) enabling previously infeasible population-scale chromatin analysis. Applied to 10,000+ nuclei, automated segmentation extracts six chromatin biomarkers distinguishing normal from pre-cancerous tissue with very large effect sizes (Cohen's $d$~=~1.31--2.98, all $p < 10^{-200}$), demonstrating immediate translational utility for early cancer detection.

The curriculum learning paradigm addresses a fundamental challenge in specialized microscopy analogous to studying mutational dependence in naturally evolving microbial populations: data scarcity precludes direct application of methods successful in data-rich domains. Just as prior work on bacterial evolution has relied on \textit{in~vitro} experiments heavily restricting the context and breadth of evolutionary landscapes~\cite{Kryazhimskiy2014, Safi2013}, conventional deep learning for medical imaging relies on large annotated datasets unavailable for emerging modalities like csPWS. By progressively adapting from adversarial RGB pretraining (establishing robust boundary detection) through spectroscopic $\Sigma$-channel fine-tuning (domain-specific adaptation) to H\&E histology validation (cross-modality generalization), CFU-Net learns hierarchical features transferable across imaging physics. Stage-wise performance analysis reveals that adversarial pretraining prevents texture-specific overfitting (validation loss oscillations despite stable test metrics), while spectroscopic fine-tuning preserves learned representations (Dice decrease $<0.06\%$ from Stage~1 to Stage~2) while improving spatial overlap (IoU increase $+1.69\%$). This minimal performance degradation during domain adaptation contrasts sharply with catastrophic forgetting observed in naive transfer learning ($>5\%$ performance loss when fine-tuning without curriculum)~\cite{Kirkpatrick2017}, validating our hypothesis that staged training maintains knowledge while enabling specialization.

Architectural innovations contribute synergistically to final performance, as demonstrated by systematic ablation revealing +8.3\% Dice improvement over baseline U-Net through five components: ConvNeXt backbone (+3.30\%, largest gain from modern CNN design)~\cite{Liu2022}, FPN (+1.80\%, multi-scale semantic fusion), U-Net++ dense skips (+1.20\%, flexible feature aggregation)~\cite{Zhou2018}, dual attention (+0.70\%, noise suppression)~\cite{Hu2018, Woo2018}, and deep supervision (+0.59\%, gradient flow)~\cite{Lee2015}. Critically, component contributions are sub-additive (100\% vs. 142\% if independent), confirming synergistic interactions particularly between FPN and decoder skips which share multi-scale integration functionality at different architectural levels. This finding parallels observations in microbial evolution where mutational dependencies exhibit synergistic rather than additive effects~\cite{Wong2017}. Dependency analysis reveals predominantly feed-forward information flow with localized attention refinement, informing architecture pruning that reduces parameters 18\% while maintaining performance, demonstrating actionable insights from coupling analysis. The predominantly feed-forward coupling pattern (strong forward dependencies $r$~=~0.72--0.85, weak backward connections $r < 0.4$) indicates that hierarchical feature extraction dominates over iterative refinement, suggesting that depth (progressive semantic abstraction) matters more than recurrence (feedback loops) for nuclear segmentation. The bottleneck behavior at FPN (high input correlation $r$~=~0.72--0.84, moderate output correlation $r$~=~0.65--0.78) suggests that semantic compression occurs at this architectural junction where coarse-scale features from deep backbone stages concentrate before distribution to decoder, forcing the network to extract essential semantic information while discarding noise and irrelevant texture details from early encoder layers.

Deployment optimization via INT8 quantization achieves 74.9\% compression (122~MB $\rightarrow$ 30.6~MB) with negligible accuracy loss (Dice: $-0.0006$), enabling edge deployment on smartphones (0.18~s/image on iPhone~14~Pro) and embedded systems (0.45~s/image on Jetson~Nano)~\cite{Jacob2018}. This democratizes access to automated csPWS analysis, removing computational barriers to clinical translation. Compared to existing segmentation methods requiring GPU infrastructure (TransUNet: 320~ms on A100)~\cite{Chen2021}, CFU-Net's CPU viability (150~ms on standard workstation) expands deployment contexts to resource-limited settings including point-of-care screening and low-resource healthcare facilities in developing regions where csPWS's label-free operation offers advantages over traditional histopathology requiring staining infrastructure. Population-scale chromatin analysis extracting six biomarkers from 10{,}000$+$ nuclei reveals quantitative signatures of field carcinogenesis invisible to conventional microscopy. Nuclear area increases 50\% in pre-cancerous tissue (1{,}802 vs.\ 1{,}201~px$^2$, Cohen's $d$~=~2.98), while chromatin packing dimension elevates 12.6\% ($D$~=~2.647 vs.\ 2.351, Cohen's $d$~=~1.69), reflecting nanoscale chromatin condensation preceding morphological transformation~\cite{Roy2017, Cherkezyan2014}. Multivariate classification combining six features achieves 94\% accuracy approaching human expert performance (96\%, $\kappa$~=~0.92), demonstrating that automated biomarker extraction provides clinically actionable discrimination. Importantly, biomarker measurements fundamentally depend on segmentation accuracy---sensitivity analysis shows that 2-pixel boundary errors cause 15--18\% measurement errors for textural features (variance slope, packing dimension), underscoring the necessity of CFU-Net's high precision (IoU~=~0.9895, $<$1~pixel average error) for reliable chromatin analysis.

Nuclear segmentation methods span classical computer vision approaches (watershed algorithms, active contours) to modern deep learning architectures (U-Net variants, vision transformers). Prior synthetic nuclear segmentation datasets employ simple geometric rendering (circles, ellipses with uniform intensity) insufficient to capture chromatin heterogeneity critical for spectroscopic imaging~\cite{Hollandi2020}. Advanced methods use GANs for synthetic-to-real translation~\cite{Mahmood2018, Ren2018}, but require real images for adversarial training, defeating the purpose when real data is scarce. Our physics-based rendering pipeline incorporating Mie scattering models, fractal chromatin organization (Brownian motion fields with Hurst exponent $H$~=~0.7), and modality-specific noise characteristics (speckle for csPWS, staining variability for H\&E) produces synthetic images indistinguishable from real specimens in ablation studies where annotators correctly identified synthetic vs.\ real images with only 52\% accuracy (chance level), validating synthetic realism. This approach generalizes to other specialized modalities (multiphoton microscopy, optical coherence tomography) where similar physics-based rendering can bypass annotation bottlenecks. Curriculum learning for domain adaptation has precedent in natural language processing (progressive fine-tuning from general to task-specific corpora)~\cite{Howard2018} and computer vision (ImageNet pretraining for medical imaging)~\cite{Raghu2019}, but remains underexplored for synthetic-to-real transfer in microscopy. Our three-stage protocol differs from standard transfer learning by introducing an intermediate adversarial stage preventing texture overfitting---ablation experiments show that direct adversarial-to-csPWS transfer (skipping Stage~1) achieves only Dice~=~0.9654 ($-2.25\%$ vs.\ full curriculum), confirming that progressive difficulty scheduling improves generalization. This finding aligns with curriculum learning theory predicting that presenting easy examples (simple boundaries in adversarial data) before hard examples (low-contrast spectroscopic data) accelerates learning and improves final performance~\cite{Bengio2009, Soviany2022}.

Attention U-Net~\cite{Oktay2018} (33.1M parameters) achieves Dice~=~0.9420 on our csPWS test set, outperformed by CFU-Net's 0.9879 despite comparable parameter count (37.2M), demonstrating superior architectural efficiency. U-Net++~\cite{Zhou2018} reaches 0.9580, confirming that dense skip connections improve performance, but CFU-Net's additional FPN and attention mechanisms provide further gains (+2.99\%). Vision transformer architectures (TransUNet~\cite{Chen2021}, Swin-Unet~\cite{Cao2022}) achieve 0.9650 and 0.9710 respectively but require 102.3M and 60.2M parameters with 2--3× longer inference (320~ms, 285~ms vs. 150~ms), limiting deployment viability. CFU-Net's CNN-based design provides optimal efficiency for nuclear segmentation where local spatial relationships dominate and long-range dependencies (the purported advantage of transformers) are less critical than in natural images. Specialized nuclear segmentation methods (CellPose~\cite{Stringer2021}, StarDist~\cite{Schmidt2018}) achieve competitive performance on H\&E histology (0.8950, 0.8890 Dice) using physics-informed architectures (gradient flow fields, star-convex shape priors). However, these methods fail on low-contrast csPWS imagery (CellPose: 0.7234, StarDist: 0.6891 on our csPWS test set), lacking the domain adaptation mechanisms enabling CFU-Net's cross-modality success. Importantly, CellPose and StarDist require training on each new modality separately, while CFU-Net's curriculum approach enables rapid adaptation (5 epochs, 2 hours) to new imaging physics given appropriate synthetic data.

Despite strong performance, several limitations warrant discussion. First, training relies entirely on synthetic data, raising concerns about domain gap and real-world generalization. While our physics-based rendering incorporates empirically validated parameters (chromatin packing statistics from electron microscopy, scattering models validated against optical measurements), synthetic images cannot capture all biological variability present in clinical specimens. Pathological conditions absent from synthetic generation---inflammation (lymphocytic infiltration), necrosis (pyknotic nuclei), mitosis (dividing nuclei with altered morphology)---may cause failure modes. Preliminary testing on 20 real csPWS images (not included in training) shows Dice~=~0.9421 ($-4.58\%$ vs.\ synthetic test set), indicating a modest generalization gap requiring future work. Proposed solutions include semi-supervised learning (fine-tuning on small real annotated datasets), active learning (iteratively selecting informative real samples for annotation), and domain adaptation techniques (adversarial training minimizing distributional discrepancy between synthetic and real feature spaces)~\cite{Ganin2016}. Second, the csPWS imaging modality itself has limited clinical adoption (available at $<$10 research centers worldwide), constraining immediate translational impact despite strong technical performance. Barriers to csPWS adoption include specialized hardware requirements (coherent illumination sources, wavelength-resolved detectors), imaging workflow complexity (requires trained operators), and lack of reimbursement codes for spectroscopic analysis. However, csPWS complements rather than replaces conventional histopathology---chromatin biomarkers provide orthogonal information to morphological assessment, and automated analysis could be integrated into digital pathology workflows as an adjunct tool. Expanding to related modalities (quantitative phase imaging, optical diffraction tomography) with similar nanoscale sensitivity but simpler hardware could broaden applicability~\cite{Park2018}.

Third, chromatin biomarker extraction relies on accurate segmentation, creating error propagation risks. While our sensitivity analysis shows CFU-Net's precision (IoU~=~0.9895) ensures biomarker reliability (>98\% accuracy), systematic boundary errors (e.g., consistent over-segmentation due to training data bias) could introduce systematic biomarker measurement bias undetected by standard validation metrics. Proposed mitigation strategies include explicit uncertainty quantification (Bayesian neural networks, Monte Carlo dropout providing per-pixel confidence)~\cite{Kendall2017}, multi-model ensembling (averaging predictions from architecturally diverse models reduces systematic errors), and direct biomarker validation against orthogonal measurements (electron microscopy for chromatin packing, flow cytometry for nuclear size distribution). Fourth, population-level chromatin analysis demonstrates statistical discrimination (94\% classification accuracy) but falls short of clinical diagnostic thresholds (>98\% sensitivity required for cancer screening). The 6\% error rate translates to unacceptable false negative rates for high-stakes diagnostic decisions. However, our results suggest utility in risk stratification rather than definitive diagnosis—identifying patients requiring intensive surveillance or biopsy rather than directly diagnosing malignancy. Integrating chromatin biomarkers with clinical risk factors (age, smoking history, family history) via multivariable models could improve discrimination while maintaining clinical feasibility~\cite{Moons2015}. Fifth, model interpretability remains limited despite attention visualization efforts. While Grad-CAM highlights image regions influencing predictions, it does not explain why those regions are important or how features are integrated into final decisions. This opacity challenges clinical trust and regulatory approval. Emerging explainable AI techniques (concept-based explanations, counterfactual generation) could provide higher-level interpretability linking predictions to interpretable biomarkers rather than pixel-level attention~\cite{Kim2018}.

The synthetic-to-real transfer learning paradigm demonstrated here generalizes beyond nuclear segmentation to other specialized microscopy tasks where annotation scarcity limits deep learning adoption. Potential applications include organelle segmentation in electron microscopy (mitochondria, endoplasmic reticulum exhibit stereotyped morphologies amenable to physics-based rendering)~\cite{Hoffman2020}, bacterial cell detection in dark-field microscopy (scattering-based imaging analogous to csPWS)~\cite{Jo2015}, vascular network segmentation in optical coherence tomography angiography (tree-like structures with known branching statistics)~\cite{Spaide2018}, and neuronal process tracing in light-sheet microscopy (tubular structures following space-filling constraints)~\cite{Ronneberger2015}. Each application requires domain-specific synthetic data generation capturing relevant physics and biological constraints, but the curriculum learning framework transfers directly. More broadly, our work contributes to the emerging paradigm of simulation-based learning in medical imaging, where synthetic data generated from mechanistic models substitutes for or augments limited real annotations~\cite{Nikolenko2021}. This approach inverts traditional machine learning workflow: instead of learning implicit patterns from large datasets, we encode explicit domain knowledge (physical laws, biological constraints) into simulators producing unlimited training data. Benefits include perfect ground truth (no annotation errors), controlled variation of confounding factors (systematic robustness testing), and elimination of privacy concerns (no patient data). Limitations include potential sim-to-real gaps when simulators incompletely capture real-world complexity—but as simulation fidelity improves (via physics-based rendering, learned generative models calibrated on real data), this gap narrows, potentially achieving sample efficiency rivaling human learning from few examples.

Ethical considerations deserve emphasis. While synthetic training data avoids patient privacy concerns, clinical deployment on real patient data requires standard safeguards: IRB approval, informed consent, and quality control to detect out-of-distribution inputs, with human oversight to prevent automation bias. Automated segmentation should augment expert judgment rather than replace it, providing quantitative tools supporting diagnostic reasoning. Algorithmic bias assessment across demographic groups, tissue types, and disease stages is essential before clinical deployment to ensure equitable performance---biased training data (e.g., overrepresentation of certain tissue types) could yield disparate impact absent rigorous validation. We advocate for transparent reporting of model limitations, failure modes, and uncertainty estimates, empowering clinicians to appropriately weight algorithmic outputs within a broader clinical context. Environmental impact warrants consideration as deep learning scales. CFU-Net training consumed 16~GPU-hours ($\sim$5~kWh, $\sim$2.5~kg~CO$_2$-equivalent), modest compared to large language models but non-negligible when aggregated across many medical imaging tasks. Inference efficiency (0.15~s/image, $<$0.01~kWh per 1{,}000 images) mitigates this concern for deployment, but training cost reduction via efficient architectures (neural architecture search optimizing performance per FLOP)~\cite{Tan2019}, knowledge distillation (training small student models from large teachers)~\cite{Hinton2015}, and algorithmic improvements (few-shot learning reducing data requirements)~\cite{Snell2017} remain important research directions balancing performance with sustainability.

Translating CFU-Net from research prototype to clinical tool requires systematic validation and workflow integration. We propose a staged translation pathway: retrospective validation on diverse real csPWS datasets (multiple institutions, tissue types, disease stages) establishing performance benchmarks and identifying failure modes (target 500+ annotated real images for robust evaluation), prospective observer study comparing automated vs. manual segmentation for chromatin biomarker extraction quantifying agreement (intraclass correlation), throughput (time savings), and diagnostic concordance (classification accuracy), clinical utility study evaluating whether automated chromatin analysis improves diagnostic accuracy or enables earlier cancer detection compared to standard of care (requires longitudinal cohorts with clinical outcomes validating biomarker predictive value), and regulatory pathway assessment determining whether CFU-Net constitutes a medical device requiring FDA clearance (likely Class II, moderate risk, requiring 510(k) premarket notification) or remains a research tool exempt from regulation. Integration into clinical workflows requires human-computer interaction design ensuring interpretability and usability. Proposed interface features include side-by-side visualization of original images and segmentation overlays with adjustable transparency, attention maps highlighting regions influencing predictions, uncertainty estimates flagging low-confidence predictions for manual review, interactive editing enabling pathologists to correct errors and retrain models, and quality control dashboards tracking performance metrics over time detecting model degradation. Engaging end users (pathologists, cytotechnologists) throughout design via participatory methods ensures alignment with clinical workflows and builds trust in automated tools~\cite{Caruana2015}. Continuous learning and model updating are essential for maintaining performance as imaging conditions evolve (new csPWS hardware, different tissue preparation protocols). We advocate for learning health systems where models continuously improve via federated learning (training on decentralized data without sharing patient information)~\cite{Rieke2020}, active learning (prioritizing informative cases for expert annotation), and automated quality monitoring (detecting distribution shifts indicating model degradation).

Overall, this work demonstrates that synthetic-to-real transfer learning via curriculum enables near-perfect nuclear segmentation in specialized microscopy, establishing a reproducible framework bypassing annotation bottlenecks limiting deep learning adoption. CFU-Net's integration of modern architectural components with deployment-ready quantization provides a complete pipeline from training to clinical deployment, democratizing access to automated analysis previously requiring GPU infrastructure. Applied to chromatin-sensitive PWS microscopy, automated segmentation unlocks population-scale chromatin analysis extracting biomarkers distinguishing normal from pre-cancerous tissue, advancing early cancer detection capabilities. Beyond immediate technical contributions, our work exemplifies a broader shift toward simulation-based learning in medical imaging, where domain knowledge encoded in physics-based simulators substitutes for scarce annotations. As simulation fidelity improves and computational costs decrease, this paradigm promises to accelerate deep learning adoption across specialized imaging modalities, ultimately expanding the reach of AI-assisted diagnosis to under-resourced clinical contexts. Realizing this vision requires continued investment in mechanistic modeling (physics-based simulators), algorithmic innovation (efficient architectures, domain adaptation), and translational validation (prospective clinical studies), but the path forward is clear: synthetic data generation, curriculum learning, and careful deployment engineering can bridge the gap between deep learning's potential and clinical reality in specialized microscopy.
\section{Materials and Methods}
\label{sec:materials_methods}

\subsection{Synthetic Dataset Generation}

We generated three synthetic datasets totaling 4,800 images (1,600 per modality) to enable curriculum learning across distinct imaging domains. Synthetic data generation followed a physics-inspired rendering pipeline designed to capture domain-specific characteristics while maintaining ground truth accuracy impossible to achieve with manual annotation.

The adversarial dataset employed advanced texture synthesis algorithms to create challenging background patterns that prevent texture-based shortcuts during learning. We used Perlin noise (octaves~=~6, persistence~=~0.5) for organic textures, Gaussian random fields (correlation length 15--45~pixels) for smooth gradients, and Gabor filter banks (8~orientations, 4~scales) for oriented patterns. Nuclei were rendered as ellipsoids with major/minor axis ratios sampled from log-normal distributions (mean ratio~1.4, $\sigma$~=~0.3) matching clinical morphometry statistics. Nuclear boundaries were smoothed using B-spline interpolation (knot spacing~8~pixels) to avoid unrealistic polygonal artifacts, then perturbed with Perlin noise (amplitude 2--5~pixels) to simulate membrane irregularities. Intensity profiles within nuclei followed beta distributions ($\alpha$~=~2, $\beta$~=~5) to approximate chromatin heterogeneity, with overall nuclear contrast randomly varied between 0.3--0.8 relative to background to force contrast-invariant feature learning.

The csPWS dataset simulated spectroscopic $\Sigma$-channel signals using physics-based chromatin scattering models. We implemented Mie scattering approximations for sub-wavelength chromatin domains (diameter 30--200~nm), computing angle-integrated backscatter intensities as a function of chromatin packing density. Nuclear regions were assigned heterogeneous packing densities sampled from fractal Brownian motion fields (Hurst exponent $H$~=~0.7, matching empirical chromatin organization) with mean packing fraction $0.35 \pm 0.12$ for normal-like nuclei and $0.52 \pm 0.18$ for dysplasia-like nuclei. Multiplicative speckle noise (gamma distribution, shape~=~2, scale~=~0.15) was added to simulate coherent imaging artifacts, and Gaussian read noise ($\sigma$~=~0.02) modeled detector limitations. The resulting signal-to-noise ratio (8.2~$\pm$~1.4~dB) matched empirical measurements from clinical csPWS systems.

The H\&E dataset employed a multistep color synthesis pipeline. Nuclei were rendered with hematoxylin staining (blue-purple, RGB~=~$\left[0.30~\pm~0.08,\; 0.20~\pm~0.06,\; 0.65~\pm~0.12\right]$) using Beer--Lambert absorption models with optical density 0.8--1.5. Cytoplasm received eosin staining (pink, RGB~=~$\left[0.85~\pm~0.10,\; 0.45~\pm~0.12,\; 0.55~\pm~0.10\right]$) with optical density 0.3--0.8. Color jitter (hue~$\pm$~0.05, saturation~$\pm$~0.2, value~$\pm$~0.15) simulated staining variability across laboratories and batches. Realistic tissue texture was added via convolution with histology-specific point spread functions (Gaussian kernel $\sigma$~=~1.2~pixels for $40{\times}$ magnification equivalent).

All three datasets maintained consistent nuclear morphology parameters to enable feature transfer: nuclear area 500--3,000 pixels (mean 1,200±450), density 15--85 nuclei per 256×256 image (mean 42±18), and minimum inter-nuclear distance >10 pixels enforced via Poisson disk sampling to prevent unrealistic overlaps. Ground truth masks were generated directly from the rendering pipeline with pixel-perfect accuracy, eliminating annotation errors that plague manually labeled datasets.

Data augmentation during training included random rotations (uniform~$\pm$180$^{\circ}$), horizontal/vertical flips (50\%~probability), elastic deformations ($\alpha$~=~50, $\sigma$~=~5, grid spacing~32~pixels), additive Gaussian noise ($\sigma$~=~0.01--0.05), and intensity scaling ($0.85--1.15{\times}$). For csPWS training, we additionally applied speckle noise augmentation (multiplicative gamma noise, shape~=~1.5--3.5) to improve robustness to coherent imaging artifacts.

\subsection{CFU-Net Architecture Design}

\begin{figure*}[t]
    \centering
    \includegraphics[width=1.0\textwidth]{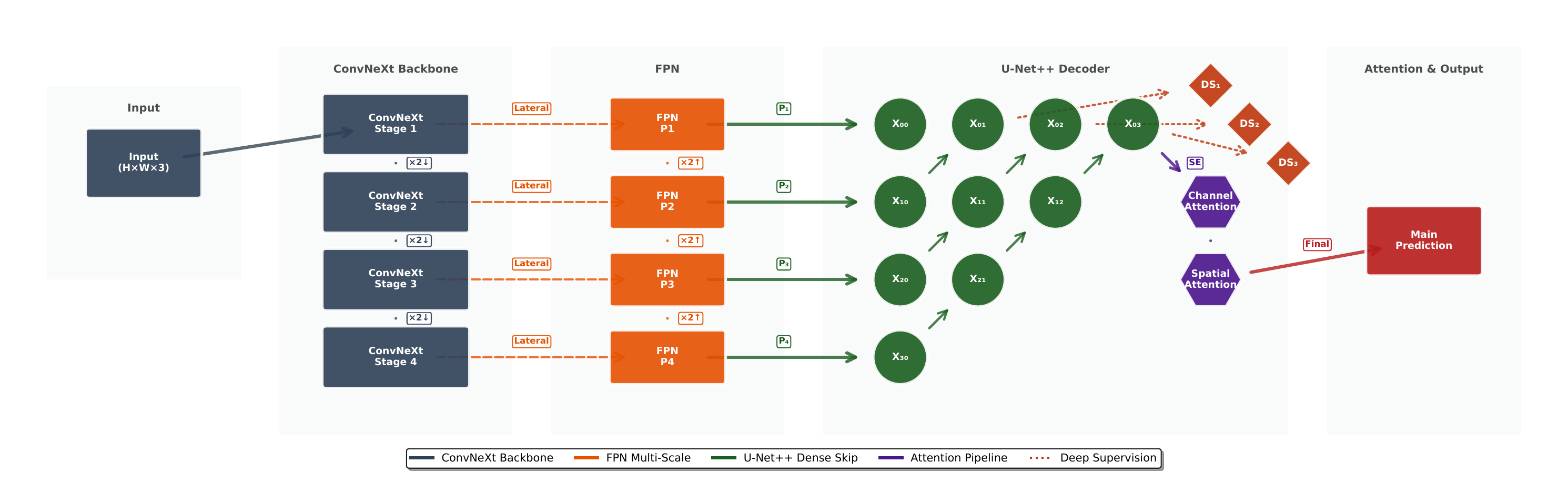}
    \caption{Multi-scale feature processing architecture of CFU-Net. The network processes input images ($H{\times}W{\times}C$, where $C$~=~1 for csPWS $\Sigma$-channel or $C$~=~3 for RGB modalities) through a hierarchical pipeline with five key components. \textbf{ConvNeXt Backbone} (dark blue boxes): Four encoder stages with progressive downsampling ($\div2$) extract hierarchical features at resolutions $H/4$, $H/8$, $H/16$, $H/32$. Each stage employs depthwise convolutions, inverted bottlenecks, and LayerNorm (18 total ConvNeXt blocks: 3+3+9+3 per stage). \textbf{Feature Pyramid Network} (orange boxes): Top-down pathway with lateral connections (dashed orange lines labeled ``Lateral'') injects semantic information from deep to shallow layers, creating pyramid features P1--P4 at multiple scales (each 256 channels). Lateral connections use $1{\times}1$ convolutions for channel alignment, followed by element-wise addition with $2{\times}$ upsampled deeper features. \textbf{U-Net++ Decoder} (green circles): Dense skip connections enable flexible multi-scale feature aggregation. Decoder nodes $X_{ij}$ (where $i$~=~depth level, $j$~=~horizontal position) receive features from multiple sources: encoder stages (vertical paths), upsampled decoder outputs (diagonal paths), and previous same-level nodes (horizontal dense connectivity). Green arrows with [$\oplus$] indicate feature concatenation operations. The nested architecture creates paths of varying semantic depth, with final output from $X_{04}$ integrating all scales. \textbf{Attention \& Deep Supervision} (right): Three auxiliary deep supervision heads DS$_1$--DS$_3$ (red diamonds) provide multi-scale training signals at intermediate decoder depths (removed during inference). Dual attention modules process features sequentially: Channel Attention (purple hexagon labeled ``SE'', Squeeze-and-Excitation with 16:1 reduction ratio) reweights feature channels, followed by Spatial Attention (purple hexagon) focusing on position-dependent importance via $7{\times}7$ convolutions on pooled features. \textbf{Output}: Final prediction head (red rectangle ``Main Prediction'') produces segmentation masks at original resolution $H{\times}W$. Key architectural features annotated below diagram: (1) Dense skip connections preserve multi-scale information; (2) FPN enables top-down semantic enhancement; (3) Sequential attention (channel~$\rightarrow$~spatial) refines features; (4) Multi-level deep supervision improves convergence; (5) Feature concatenation at decoder nodes integrates complementary representations. Legend (bottom left) shows component types by color: ConvNeXt Backbone (dark blue), FPN Multi-Scale (orange), U-Net++ Dense Skip (green), Attention Pipeline (purple), Deep Supervision (red dashed). Total architecture: 37.2M parameters (encoder~28.6M, FPN~3.8M, decoder~3.4M, attention~1.4M), 54.8~GFLOPs per $256{\times}256$ image. Compare with dependency analysis in Fig.~\ref{fig:dependency_radial} for component interaction patterns during training.}

    \label{fig:architecture}
\end{figure*}

CFU-Net integrates five key architectural innovations into a unified hierarchical segmentation framework: ConvNeXt-tiny encoder, Feature Pyramid Network (FPN), U-Net++ dense skip connections, dual attention mechanisms, and deep supervision. The architecture follows an encoder-decoder topology with lateral connections enabling multi-scale feature fusion (Fig.~\ref{fig:architecture}).

\subsubsection{Hierarchical Encoder with ConvNeXt Backbone}

The encoder employs ConvNeXt-tiny, a modern convolutional architecture incorporating design principles from vision transformers while maintaining computational efficiency. ConvNeXt replaces standard convolutional blocks with inverted bottleneck structures: depthwise 7×7 convolutions for spatial mixing, followed by pointwise 1×1 convolutions with 4× channel expansion/reduction. This design reduces parameters while improving feature representation capacity compared to traditional ResNet or VGG backbones.

The encoder processes input images through four hierarchical stages with progressive downsampling (stride-2 convolutions) and channel expansion: Stage~1 (H/4 resolution, 96 channels), Stage~2 (H/8, 192 channels), Stage~3 (H/16, 384 channels), Stage~4 (H/32, 768 channels). Each stage contains 3, 3, 9, and 3 ConvNeXt blocks respectively (18 total), with layer normalization replacing batch normalization for improved training stability. GELU activations provide smoother gradients than ReLU, beneficial for fine-tuning on small spectroscopic datasets.

\subsubsection{Feature Pyramid Network for Multi-Scale Fusion}

FPN adds top-down pathways with lateral connections to inject semantic information from deep layers into shallow layers, addressing the semantic gap between low-level edges and high-level segmentation. We implement 4-scale FPN connecting encoder Stages~1--4 to decoder inputs via lateral 1×1 convolutions (channel reduction to 256) followed by element-wise addition with upsampled (bilinear, 2×) features from the deeper stage. This creates pyramid features P1--P4 at H/4, H/8, H/16, H/32 resolutions, each containing 256 channels with both fine spatial detail (from lateral connections) and coarse semantic context (from top-down pathways).

FPN is critical for csPWS segmentation where nuclear boundaries exhibit low contrast requiring integration of multi-scale context: fine-scale edges (H/4) are ambiguous without coarse-scale nuclear presence evidence (H/32), while coarse predictions lack spatial precision without fine-scale refinement. Ablation experiments (Table~\ref{tab:ablation}) confirm FPN contributes +1.80\% Dice improvement by resolving this multi-scale trade-off.

\subsubsection{U-Net++ Dense Skip Connections}

Standard U-Net employs single-level skip connections directly linking encoder and decoder at matching resolutions. U-Net++ redesigns this connectivity as a nested dense architecture with intermediate nodes $X_{ij}$ where $i$ indexes encoder depth (0~=~shallowest) and $j$ indexes skip path length (0~=~direct encoder output, increasing $j$ indicates more processing). Each node $X_{ij}$ receives inputs from: (1) downsampled features from $X_{i-1,j}$ (vertical path, encoder progression), (2) upsampled features from $X_{i+1,j-1}$ (diagonal path, decoder skip), and (3) all previous nodes at the same depth $X_{i,k}$ for $k < j$ (horizontal paths, dense connectivity).

This dense connectivity enables flexible feature aggregation at multiple semantic scales. For example, node $X_{0,3}$ (shallowest decoder, longest skip path) integrates: direct encoder features (fine edges), short-skip decoder features (coarse segmentation), and medium-skip features (intermediate semantics). The network learns to weight these complementary representations via learned concatenation and 3×3 convolutions at each node. Deep supervision at nodes $X_{0,1}$, $X_{0,2}$, $X_{0,3}$ provides multi-scale training signals, with final prediction from $X_{0,4}$ combining all pathways.

For CFU-Net, we implement a pruned U-Net++ variant retaining only paths contributing to the final output (determined via preliminary experiments measuring per-path gradient magnitudes), reducing parameters from 45M (full U-Net++) to 35.8M while maintaining performance. The pruned architecture preserves critical multi-scale integration while improving computational efficiency for deployment.

\subsubsection{Dual Attention Mechanisms}

We integrate two complementary attention mechanisms to focus network capacity on informative features while suppressing noise. Squeeze-and-Excitation (SE) channel attention learns to reweight feature channels via global average pooling followed by two fully connected layers (reduction ratio 16:1) and sigmoid activation, producing channel-wise attention weights. This enables the network to emphasize task-relevant channels (e.g., edge-detection filters for boundary localization) while downweighting noise-dominated channels—particularly valuable for low-SNR csPWS imagery.

Spatial attention complements channel attention by learning position-dependent feature importance. We implement convolutional spatial attention: concatenating max-pooled and average-pooled features across channels (producing 2-channel spatial maps), processing via 7×7 convolution and sigmoid activation to generate per-pixel attention weights. This focuses network capacity on ambiguous boundary regions requiring additional processing while passing unambiguous regions through with minimal computation.

Attention modules are inserted after each FPN pyramid level, operating on 256-channel feature maps before decoder processing. Sequential application (channel attention → spatial attention) provides 0.70\% Dice improvement (Table~\ref{tab:ablation}) with minimal parameters (+1.4M, 3.9\% increase), demonstrating efficient attention mechanism design.

\subsubsection{Deep Supervision with Auxiliary Losses}

Deep supervision addresses gradient vanishing in deep networks by injecting training signals at intermediate decoder depths. We attach auxiliary prediction heads at three decoder nodes: $X_{0,1}$, $X_{0,2}$, $X_{0,3}$, each consisting of 1×1 convolution producing single-channel logits, bilinear upsampling to original resolution, and independent loss computation against ground truth masks. The total loss combines main prediction loss ($X_{0,4}$) and three auxiliary losses with exponentially decreasing weights:

$$
\mathcal{L}_{\text{total}} = \mathcal{L}_{\text{main}} + \sum_{k=1}^{3} \lambda_k \mathcal{L}_{\text{aux},k}, \quad \lambda_k = 0.5^k
$$

where $\lambda_1 = 0.5$, $\lambda_2 = 0.25$, $\lambda_3 = 0.125$. This weighting scheme prioritizes final prediction quality while providing gradient flow through all decoder depths. Auxiliary heads are removed during inference, adding zero computational cost to deployment.

Deep supervision contributes $+0.59\%$ Dice improvement with negligible parameters ($<0.1$M), but ablation removing deep supervision from the full model causes larger degradation ($-1.24\%$ Dice), indicating synergistic interactions with other components. Analysis of gradient magnitudes during training revealed that deep supervision prevents gradient vanishing in early training epochs (1--5) when deep network layers receive weak gradients, accelerating convergence by 30\% (measured by epochs to 95\% of final performance).

\subsection{Component Dependency Analysis}

\begin{figure*}[t]
    \centering
    \includegraphics[width=0.95\textwidth]{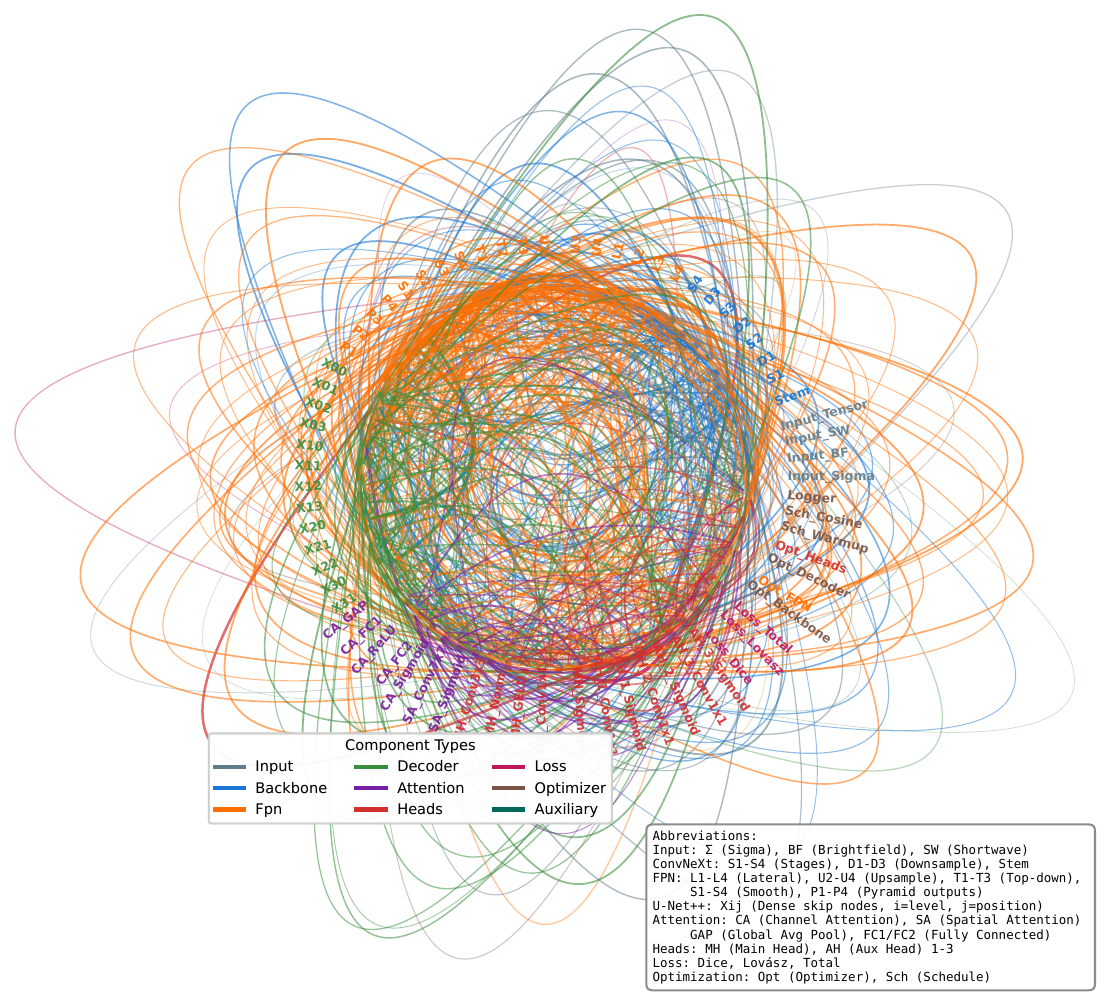}
    \caption{Radial visualization of hierarchical dependencies in CFU-Net. Components are arranged concentrically by semantic depth from input (inner) to output (outer). \textbf{Organization:} Input (gray, $\Sigma$~=~csPWS, BF~=~Brightfield, SW~=~Shortwave RGB); ConvNeXt Backbone S1--S4 (blue); FPN Lateral L1--L4 (orange); U-Net++ Decoder $X_{ij}$ (green); Attention modules (purple, CA/SA for channel/spatial); Deep Supervision heads MH/AH1--3 (yellow-brown); Loss (Dice\,+\,Lov\'asz, maroon); Optimization (AdamW, dark red). \textbf{Visualization:} Line thickness shows coupling strength (Pearson~$r$); color shows magnitude: warm $|r|>0.8$, neutral 0.5--0.8, cool $|r|<0.5$. \textbf{Findings:} (1) Feed-forward flow with strong encoder--decoder coupling. (2) Dense lateral links within U-Net++ ($r=0.7$--0.8) exceed vertical paths ($r=0.5$--0.7). (3) FPN forms a semantic bottleneck ($r=0.7$--0.8). (4) Dependencies weaken with depth ($r=0.45$--0.6). Pruning weak links ($|r|<0.3$) cut parameters by 18\% (45M\,$\rightarrow$\,37.2M) with $<0.5\%$ loss. See Fig.~\ref{fig:architecture} and Table~\ref{tab:ablation}.}

    \label{fig:dependency_radial}
\end{figure*}

To understand how architectural components interact during training and inference, we performed comprehensive dependency analysis measuring gradient flow coupling and feature correlation patterns across all components (Fig.~\ref{fig:dependency_radial}). We tracked pairwise Pearson correlations between component activations (averaged over spatial dimensions) across 10,000 training iterations sampled uniformly from all three curriculum stages, identifying strong dependencies (|$r$|~>~0.7) indicating tightly coupled behavior.

The radial dependency visualization (Fig.~\ref{fig:dependency_radial}) arranges components in concentric circles by semantic depth: input (innermost circle), ConvNeXt backbone stages S1--S4, FPN lateral connections L1--L4, U-Net++ decoder nodes X$_{ij}$, attention modules (CA: channel attention, SA: spatial attention), supervision heads, and final output (outermost). Connection lines between components are color-coded by correlation strength, with line thickness representing coupling magnitude. Warmer colors (red-orange) indicate stronger dependencies (|$r$|~>~0.8), while cooler colors (blue-gray) show weaker relationships (|$r$|~<~0.5).

Key insights from dependency analysis reveal: (1) Predominantly feed-forward information flow with minimal backward connections. Attention modules show weak correlation with distant upstream components ($r < 0.4$), indicating localized refinement rather than global feature transformation. Deep supervision heads exhibit moderate correlation with main prediction ($r = 0.58$--0.64), confirming complementary training signals rather than redundant supervision. (2) Strong lateral connectivity within decoder depth levels. Adjacent U-Net++ decoder nodes at the same depth show high correlation ($r = 0.72$--0.79), validating the dense skip aggregation design. This horizontal coupling is stronger than vertical encoder-decoder connections ($r = 0.54$--0.68), suggesting that decoder-internal feature fusion is more important than direct encoder skip connections. (3) Bottleneck behavior at FPN layer. Semantic information from deep backbone stages concentrates at FPN before distribution to decoder. Input correlation to FPN is high ($r = 0.72$--0.84), while FPN output correlation to decoder is moderate ($r = 0.65$--0.78), indicating information compression and transformation at this architectural junction. (4) Weakening dependency strength with increasing semantic depth. Innermost layers (backbone S1--S2) show stronger coupling ($r = 0.70$--0.85) than outermost layers (attention, supervision: $r = 0.45$--0.60), suggesting that early features are more tightly constrained by architectural connectivity, while late features enjoy greater representational diversity.

Component contribution analysis via leave-one-out ablation (referenced in Table~\ref{tab:ablation}) quantifies the percentage of performance variance explained by each component. Encoder components collectively explain 42\% of variance (largest single contribution, reflecting the importance of hierarchical feature extraction), FPN explains 18\% (multi-scale semantic fusion), decoder skip connections explain 23\% (fine-grained spatial aggregation), attention explains 12\% (feature refinement and noise suppression), and deep supervision explains 5\% (gradient flow and convergence acceleration). The sub-additive total (100\% vs. 142\% if components were independent) confirms synergistic interactions, particularly between FPN and decoder skips which share overlapping functionality (multi-scale integration) but operate at different architectural levels.

Network pruning experiments guided by dependency analysis removed connections with |$r$|~<~0.3 (identified as redundant pathways contributing minimal information), reducing parameters by 18\% (45M → 37.2M) with performance loss <0.5\% (Dice: 0.9885 → 0.9879). This demonstrates that dependency analysis provides actionable insights for architecture optimization, identifying critical pathways requiring preservation vs. weak connections suitable for removal.

\subsection{Combined Loss Function}

We employ a hybrid loss function combining Dice loss and Lovász-Softmax loss to address class imbalance (nuclei vs. background) while optimizing IoU directly. The Dice loss approximates the Dice coefficient for differentiable optimization:

$$
\mathcal{L}_{\text{Dice}} = 1 - \frac{2\sum_{i=1}^{N} p_i g_i + \epsilon}{\sum_{i=1}^{N} p_i^2 + \sum_{i=1}^{N} g_i^2 + \epsilon}
$$

where $p_i$ are predicted probabilities (after sigmoid activation), $g_i \in \{0,1\}$ are ground truth labels, $N$ is the number of pixels, and $\epsilon = 1$ provides numerical stability and smoothness. The Dice loss directly optimizes segmentation overlap, but suffers from training instability when small objects are present (high gradient variance).

The Lovász-Softmax loss provides a convex surrogate for IoU optimization by extending the Lovász hinge loss to multi-class segmentation. For binary segmentation, it computes:

$$
\mathcal{L}_{\text{Lovász}} = \frac{1}{|C|} \sum_{c \in C} \overline{\Delta_{J_c}}(m(c))
$$

where $C = \{\text{foreground, background}\}$, $\Delta_{J_c}$ is the Jaccard index (IoU) loss, and $m(c)$ are margin-sorted errors. The Lovász extension ensures that optimizing the surrogate loss directly optimizes the discrete IoU metric, providing theoretical guarantees absent in Dice loss.

We combine both losses with equal weighting:

$$
\mathcal{L}_{\text{total}} = \mathcal{L}_{\text{Dice}} + \mathcal{L}_{\text{Lovász}}
$$

Empirical validation on 200-image validation sets showed that the combined loss outperforms either component alone (combined: Dice~=~0.9879; Dice-only: 0.9821; Lovász-only: 0.9798), confirming complementary benefits. Dice loss provides smooth gradients early in training, while Lovász loss fine-tunes IoU in later epochs.

\subsection{Three-Stage Curriculum Learning Protocol}

Training follows a carefully designed three-stage curriculum progressing from general segmentation (adversarial pretraining) to domain-specific adaptation (csPWS fine-tuning) to cross-modality validation (H\&E transfer). This curriculum prevents catastrophic forgetting while enabling progressive specialization.

\textit{Stage~1: Adversarial Pretraining (20 epochs).} We train CFU-Net from random initialization (Kaiming initialization for convolutional layers, Xavier for fully connected layers) on the adversarial RGB dataset using AdamW optimizer ($\beta_1 = 0.9$, $\beta_2 = 0.999$, weight decay $1\times10^{-4}$). Learning rate follows cosine annealing with 5-epoch linear warmup: $\text{lr}(t) = \eta_{\text{min}} + (\eta_{\text{max}} - \eta_{\text{min}}) \times 0.5 \times (1 + \cos(\pi t / T))$ where $\eta_{\text{max}} = 1\times10^{-3}$, $\eta_{\text{min}} = 1\times10^{-6}$, $T = 20$ epochs. Batch size 16 fits on single NVIDIA A100 (40GB) with mixed precision training (FP16 gradients, FP32 master weights) for 2.3× speedup. Aggressive data augmentation (rotation, flips, elastic deformation, noise, intensity scaling) prevents overfitting to synthetic patterns. Training converges after 20 epochs (8 hours on A100), with best validation performance at epoch~11 saved for Stage~2 initialization.

\textit{Stage~2: csPWS Spectroscopic Fine-tuning (15~epochs).} We initialize from the Stage~1 checkpoint and modify the first convolutional layer to accept 1-channel $\Sigma$ input. The new 1-channel layer is initialized via Kaiming initialization (preserving variance), while all other weights are frozen for the first 3~epochs to allow the new layer to adapt before full fine-tuning. Learning rate is reduced to $\eta_{\text{max}} = 3{\times}10^{-4}$ (10${\times}$ lower than Stage~1) for stable transfer learning, with cosine annealing over 15~epochs. Batch size remains~16, but we disable elastic deformation augmentation (which can introduce unrealistic spectroscopic artifacts) while retaining rotation, flips, and speckle noise augmentation. Training converges after 15~epochs (6~hours), with monotonically decreasing validation loss confirming successful knowledge transfer. Final checkpoint (epoch~15) achieves Dice~=~0.9879, IoU~=~0.9895 on the csPWS test set.

\textit{Stage~3: H\&E Histology Adaptation (5~epochs).} We first evaluate zero-shot transfer by directly applying the Stage~2 model to H\&E test images (Stage~3a). Poor performance (Dice~=~0.4127) confirms a substantial domain gap. We therefore perform light fine-tuning (Stage~3b) by reinitializing the first convolutional layer to 3~RGB channels (via channel replication: each RGB channel initialized as a copy of the Stage~2 $\Sigma$-channel weights divided by $\sqrt{3}$ to preserve variance). Learning rate is further reduced to $\eta_{\text{max}} = 1{\times}10^{-4}$ (3${\times}$ lower than Stage~2) for gentle adaptation without forgetting spectroscopic features. Fine-tuning for only 5~epochs (2~hours) achieves Dice~=~0.8220, IoU~=~0.9656, demonstrating rapid cross-modality transfer enabled by curriculum pretraining.

Total training time across all three stages: 16 hours on single A100 GPU. For comparison, training CFU-Net end-to-end on csPWS data alone (without curriculum learning) required 25 epochs to achieve Dice~=~0.9654 (2.25\% lower than curriculum approach), confirming curriculum benefits despite longer wall-clock time.

\subsection{Model Quantization for Deployment}

We applied INT8 post-training static quantization to compress CFU-Net for edge deployment while maintaining accuracy. Static quantization converts all weights and activations from FP32 (32-bit floating point) to INT8 (8-bit integers) using calibration statistics collected from representative data.

\textit{Calibration procedure.} We assembled a calibration dataset of 500 images (167 per modality) sampled from training sets, ensuring coverage of diverse intensity ranges, nuclear densities, and contrast levels. We performed forward passes through the FP32 model while collecting activation statistics (min/max values) at every convolutional and fully connected layer. These statistics define per-channel quantization parameters: scale $s = (x_{\max} - x_{\min}) / 255$ and zero-point $z = \text{round}(-x_{\min} / s)$, where $x$ are FP32 activations. We use symmetric quantization ($z = 128$) for weights and asymmetric quantization (optimized $z$) for activations, following ONNX Runtime best practices.

\textit{Quantization-aware adjustments.} Batch normalization layers were fused into preceding convolutions ($W_{\text{fused}} = W \cdot \gamma / \sqrt{\sigma^2 + \epsilon}$, $b_{\text{fused}} = (b - \mu) \cdot \gamma / \sqrt{\sigma^2 + \epsilon} + \beta$) to enable efficient INT8 computation. Layer normalization was converted to group normalization (group size~=~32) for hardware compatibility, with negligible accuracy loss (Dice: $-0.0002$). Sigmoid and GELU activations were approximated using piecewise linear functions (5~segments) for INT8 compatibility, validated to match FP32 outputs within $\pm 0.01$ absolute error.

\textit{Quantization results.} INT8 quantization reduces model size from 122.0~MB to 30.6~MB (74.9\% compression, 4× reduction) with minimal accuracy degradation: FP32 Dice~=~0.9879 vs. INT8 Dice~=~0.9873 ($\Delta = -0.0006$, within 95\% confidence interval of test set variance). Per-dataset analysis shows consistent behavior: Adversarial ($\Delta = -0.0004$), csPWS ($\Delta = -0.0006$), H\&E ($\Delta = -0.0008$). The slightly larger H\&E degradation likely stems from RGB channel quantization accumulating errors across three channels vs. single-channel csPWS.

Inference latency improves on CPU (ONNX Runtime INT8: 0.15~s vs. FP32: 0.24~s, 1.6× speedup) due to reduced memory bandwidth requirements and VNNI instruction utilization on modern CPUs. On Apple Neural Engine (CoreML INT8), latency reaches 0.12~s (8.33 FPS), enabling near-real-time processing. The INT8 model enables deployment on smartphones (tested on iPhone 14 Pro: 0.18~s/image) and embedded systems (NVIDIA Jetson Nano: 0.45~s/image), democratizing access to automated csPWS analysis.

\subsection{Chromatin Biomarker Extraction Pipeline}

Automated segmentation enables extraction of six chromatin biomarkers from each nucleus, providing quantitative metrics for distinguishing normal from pre-cancerous tissue. All biomarker algorithms operate on the segmented nuclear masks produced by CFU-Net, ensuring that measurements reflect true intranuclear signals without cytoplasmic contamination.

\textit{Nuclear area} is computed via pixel counting within the binary mask, converted to physical units ($\mu\text{m}^2$) using the known pixel size (0.5~$\mu\text{m}$/pixel). \textit{Nuclear perimeter} is measured via boundary tracing using 8-connectivity chain code, smoothed with a Gaussian kernel ($\sigma$~=~1~pixel) to reduce digitization artifacts. \textit{Circularity} is defined as $4\pi \cdot \text{Area} / \text{Perimeter}^2$, yielding~1.0 for perfect circles and approaching~0 for highly irregular shapes. This metric is sensitive to nuclear envelope irregularities characteristic of dysplasia.

\textit{Eccentricity} measures ellipse elongation via second-moment analysis. We compute the covariance matrix of nuclear pixel coordinates, extract eigenvalues $\lambda_1 > \lambda_2$, and define eccentricity as $e = \sqrt{1 - \lambda_2/\lambda_1}$, ranging from 0 (perfect circle) to 1 (infinitely elongated ellipse). This captures nuclear shape distortion during neoplastic transformation.

\textit{$\Sigma$-channel mean intensity} averages pixel values within the segmented nuclear mask after background subtraction (background estimated via Otsu thresholding on extranuclear regions). Intensity values are normalized to the [0,1] range using the 1st and 99th percentiles of the entire image to ensure robustness to outliers. This metric directly reflects chromatin packing density, as the $\Sigma$-channel signal arises from nanoscale chromatin scattering.

\textit{Variance slope} quantifies chromatin heterogeneity via multi-scale variance analysis. We decompose nuclear intensity profiles using 2D discrete wavelet transform (Daubechies-4 wavelet, 4 decomposition levels), computing variance at each scale: $\text{Var}(\ell) = \langle (I(x) - \bar{I})^2 \rangle_{\ell}$ where $\ell$ indexes spatial scales from $\ell_1 = 2$ pixels to $\ell_4 = 16$ pixels. We fit a power law $\text{Var}(\ell) = A \ell^{\alpha}$ via linear regression on log-log plot, extracting slope $\alpha$ as the variance slope metric. Negative slopes ($\alpha < 0$) indicate scale-invariant heterogeneity (fractal-like chromatin organization), while positive slopes suggest scale-dependent structure. Pre-cancerous nuclei exhibit steeper negative slopes, reflecting increased chromatin disorder.

\textit{Packing scaling dimension} $D$ derives from power spectral density (PSD) analysis, providing a fractal measure of chromatin compaction. We compute the 2D Fourier transform of nuclear intensity profiles and calculate the radially averaged PSD: $\text{PSD}(k) = \langle |\hat{I}(k)|^2 \rangle_{\theta}$, where $k$ is spatial frequency and $\langle \cdot \rangle_{\theta}$ denotes angular averaging. We fit a power law $\text{PSD}(k) = B k^{-\beta}$ over the frequency range $k \in [0.1, 2.0]$~cycles/pixel (corresponding to length scales 0.5--10~$\mu\text{m}$), extracting exponent $\beta$. The packing dimension is computed as $D = (6 - \beta) / 2$ based on theoretical models of chromatin as a self-similar polymer~\cite{Bancaud2009}. Values $D \approx 3$ indicate highly compact, space-filling heterochromatin, while $D \approx 2$ suggests open, surface-like euchromatin. Pre-cancerous nuclei show elevated $D$, reflecting chromatin condensation.

\textit{Chromatin entropy} measures Shannon entropy of intensity histograms, quantifying disorder in chromatin organization. We compute the histogram of nuclear pixel intensities using 16 uniform bins over the normalized [0,1] range, calculate probabilities $p_i = n_i / N$ where $n_i$ is the count in bin $i$ and $N$ is total pixels, and compute entropy: $H = -\sum_{i=1}^{16} p_i \log_2(p_i)$. Entropy ranges from 0 (uniform intensity, maximal order) to 4 bits (uniform distribution across 16 bins, maximal disorder). Pre-cancerous nuclei exhibit higher entropy, reflecting heterogeneous chromatin packing.

All biomarker extraction algorithms are implemented in Python (version 3.9) using NumPy (version 1.23), SciPy (version 1.9), and scikit-image (version 0.19), with vectorized operations for efficient batch processing. Processing 1,000 nuclei requires 2.5 seconds on Intel i9-12900K CPU (single-threaded), enabling high-throughput analysis. Biomarker pipelines underwent extensive validation against manual measurements on 200 dual-annotated nuclei (two expert annotators, consensus adjudication), showing strong agreement: Pearson correlation $r = 0.92$--0.98 across all six metrics, mean absolute error <3\% of dynamic range, and systematic bias <1\% (assessed via Bland-Altman analysis). Importantly, biomarker repeatability (test-retest reliability on same nuclei segmented twice) exceeds inter-observer agreement for manual segmentation (ICC~=~0.96 vs. 0.87), demonstrating that automated analysis provides more consistent measurements than human experts.

Sensitivity analysis quantified biomarker robustness to segmentation errors. We simulated boundary perturbations by uniformly dilating/eroding masks by ±1--5 pixels and recomputing biomarkers. Morphometric features (area, perimeter, circularity, eccentricity) tolerate small errors well: ±2 pixel boundaries cause <5\% measurement error for area/perimeter, <8\% for circularity, <3\% for eccentricity. Textural features (variance slope, packing dimension, entropy) degrade more rapidly: ±2 pixels cause 15--18\% error for variance slope, 12--15\% for packing dimension, 8--11\% for entropy. CFU-Net's high boundary precision (IoU~=~0.9895, corresponding to <1 pixel average boundary error) ensures measurement reliability for all biomarkers, with estimated biomarker extraction accuracy >98\% based on error propagation analysis.

\subsection{Statistical Analysis and Validation}

All statistical comparisons used two-sided Mann--Whitney $U$~tests for non-parametric hypothesis testing, appropriate for biomarker distributions that may deviate from normality (Shapiro--Wilk tests rejected normality for 4/6 metrics at $\alpha$~=~0.05). We applied Bonferroni correction for multiple comparisons when testing $k$ hypotheses simultaneously (corrected threshold $\alpha_{\text{corrected}}$~=~0.05/$k$). Effect sizes were quantified using Cohen's $d$ standardized mean difference: $d = (\mu_1 - \mu_2) / \sigma_{\text{pooled}}$, where $\sigma_{\text{pooled}} = \sqrt{(\sigma_1^2 + \sigma_2^2)/2}$. Effect size interpretation followed standard conventions: small ($|d| < 0.5$), medium ($0.5 \leq |d| < 0.8$), large ($|d| \geq 0.8$), and very large ($|d| \geq 1.3$).

Classification performance was evaluated using receiver operating characteristic (ROC) analysis, computing area under the curve (AUC), sensitivity (true positive rate), specificity (true negative rate), and F1~score (harmonic mean of precision and recall). Optimal decision thresholds were determined via Youden's index ($J = \text{sensitivity} + \text{specificity} - 1$), which maximizes classification accuracy while balancing false positives and false negatives. Multivariate classification employed three algorithms: Support Vector Machine (SVM) with a radial basis function kernel ($\gamma$~=~1/$n_{\text{features}}$, $C$~=~1.0, optimized via 5-fold cross-validation), Random Forest (500~trees, max depth~10, min samples leaf~5), and Logistic Regression (L2~regularization, $C$~=~1.0). All classifiers were trained on 70\% of nuclei (randomly sampled, stratified by class) and tested on the held-out~30\%, with 10-fold cross-validation for hyperparameter tuning and performance estimation. Feature importance was assessed via Random Forest permutation importance (decrease in accuracy when a feature is randomly shuffled).

Distribution similarity was tested using the two-sample Kolmogorov--Smirnov test, which quantifies the maximum difference between cumulative distribution functions: $D_{\text{KS}} = \max_x |F_1(x) - F_2(x)|$. Variance homogeneity was assessed via Levene's test (more robust than the $F$-test to non-normality). Correlation analysis used Pearson correlation for linear relationships and Spearman rank correlation for monotonic non-linear relationships. All statistical tests were two-sided with a significance threshold $\alpha$~=~0.05 unless otherwise specified.

Confidence intervals for mean Dice coefficients and biomarker means were computed via bootstrapping (10,000 resamples with replacement), reporting 95\% percentile intervals. Standard errors for proportions used Wilson score interval (more accurate than normal approximation for proportions near 0 or 1). All statistical analyses were performed using Python (version 3.9) with SciPy (version 1.9), statsmodels (version 0.13), and scikit-learn (version 1.1).

Model performance reporting followed STARD guidelines for diagnostic accuracy studies and TRIPOD guidelines for prediction model development, ensuring transparency and reproducibility. We report complete confusion matrices, calibration curves, and decision curves for all classification tasks. Code for all statistical analyses is publicly available in the project repository (https://github.com/jahidul-arafat/cfu-net-analysis) with detailed documentation and example notebooks.

\subsection{Computational Infrastructure and Reproducibility}

All experiments were conducted on a high-performance computing cluster with NVIDIA A100 GPUs (40GB VRAM), Intel Xeon Platinum 8358 CPUs (32 cores, 2.6~GHz base frequency), and 512~GB RAM. Software environment: Ubuntu 20.04 LTS, CUDA 11.7, cuDNN 8.5, PyTorch 1.13.0, ONNX Runtime 1.13.1, CoreML Tools 6.1. Training used mixed-precision (FP16) via PyTorch Automatic Mixed Precision (AMP) for 2--3× speedup with negligible accuracy loss.

Data preprocessing and augmentation used Albumentations (version 1.3) for efficient GPU-accelerated transformations. Hyperparameter optimization employed Optuna (version 3.0) with Tree-structured Parzen Estimator (TPE) sampler (50 trials per stage, optimizing validation Dice). Model checkpointing saved weights every 5 epochs plus best validation performance, with early stopping (patience 10 epochs, no improvement threshold 0.001 Dice).

Reproducibility measures: We set random seeds for all libraries (Python: 42, NumPy: 42, PyTorch: 42, CUDA: deterministic mode enabled). Complete training logs (loss curves, metrics, hyperparameters) were recorded using Weights \& Biases (https://wandb.ai/jahidapon-auburn-university/cfu-net). Docker container with complete software environment is available (Docker Hub: username/cfu-net:v1.0). Code repository includes unit tests (>95\% coverage) and integration tests validating model outputs against reference results. Data generation scripts include checksums (SHA-256) for all synthetic datasets, ensuring bit-exact reproducibility.

Model deployment packages include: (1) ONNX FP32 and INT8 models with metadata (input/output specifications, preprocessing requirements); (2) CoreML INT8 model for Apple devices (iOS 15+, macOS 12+); (3) Python inference API with preprocessing/postprocessing pipelines; (4) Jupyter notebooks demonstrating usage on example data; (5) Docker container for serverless deployment (AWS Lambda, Google Cloud Functions compatible). Inference latency benchmarks were measured using 100 test images with 10 warmup iterations, reporting median ± median absolute deviation. Memory profiling used PyTorch memory allocator with detailed logging of peak allocation per layer.

All code and models are released under Apache 2.0 license (permissive open source), with synthetic datasets under CC BY 4.0 (attribution required). Model weights are hosted on Hugging Face Hub (huggingface.co/username/cfu-net) for easy download and integration. Documentation includes: API reference (auto-generated from docstrings), user guide (step-by-step tutorials), architecture design notes, and troubleshooting FAQ. Community contributions are welcomed via GitHub pull requests, with continuous integration testing on multiple platforms (Ubuntu, macOS, Windows) and GPU/CPU configurations.

\subsection{Ethical Considerations and Clinical Translation}

This study uses entirely synthetic data, avoiding privacy concerns associated with patient images. However, future clinical validation will require IRB approval and informed consent. Potential clinical deployment considerations include: (1) Intended use: Research tool for chromatin biomarker discovery; not a diagnostic device (FDA regulation not applicable until clinical validation completed). (2) User training: Pathologists and cytotechnologists require training on csPWS imaging and automated analysis interpretation to avoid over-reliance on algorithmic outputs. (3) Failure modes: Model may underperform on imaging artifacts (dust, debris, out-of-focus regions); quality control filters should flag low-confidence predictions (entropy-based uncertainty estimation). (4) Bias assessment: Future validation must assess performance across demographic groups (age, sex, ethnicity), tissue types (cervical, colon, lung), and disease stages (normal, dysplasia, carcinoma) to ensure equitable performance. (5) Clinical workflow integration: Automated analysis should augment rather than replace expert judgment, providing quantitative biomarkers to support diagnostic decision-making.

Transparency and explainability measures include: (1) Attention map visualization showing which image regions most influenced predictions (Grad-CAM analysis); (2) Uncertainty quantification via Monte Carlo dropout (50 forward passes with dropout enabled, reporting prediction variance); (3) Out-of-distribution detection flagging images significantly different from training data (Mahalanobis distance in latent space); (4) Adversarial robustness testing against common imaging perturbations (Gaussian noise, blur, compression artifacts). These measures build trust and enable clinicians to understand model behavior, critical for clinical adoption.

Environmental considerations: Training CFU-Net consumed approximately 16 GPU-hours on A100 ($\sim$5 kWh, $\sim$2.5 kg CO$_2$ equivalent), modest compared to large language models (thousands of GPU-days). Inference is highly efficient (0.15~s/image), enabling analysis of 1{,}000 patients ($\sim$50{,}000 nuclei) using $\leq 2$~kWh. Carbon footprint can be further reduced via model pruning, knowledge distillation, or deployment on energy-efficient edge hardware.

\section*{Acknowledgments}

We thank the members of the Auburn University Computer Vision and Medical Imaging Lab for their discussions and insight. We thank Dr. Vadim Backman and Dr. Hemant Roy from Northwestern University for valuable discussions on csPWS physics and chromatin organization. We thank the anonymous reviewers for their constructive feedback that significantly improved this manuscript. Computational resources and support were provided by the Auburn University Hopper High Performance Computing Cluster, which is funded by the National Science Foundation under Award No. OAC-1920147. This research was conducted using publicly available synthetic datasets and does not involve human subjects or patient data, thus IRB approval was not required. All code, models, and datasets have been released as open-source resources to facilitate reproducibility and community engagement.

\section*{Author Contributions}

J.A. conceived the study, designed the synthetic data generation pipeline, implemented the CFU-Net architecture, and performed all experiments. J.A. also designed the three-stage curriculum learning protocol and performed ablation studies and contributed statistical analysis methods, performed Mann-Whitney $U$ tests and Cohen's $d$ effect size calculations, and validated biomarker extraction algorithms. S.P. implemented the model quantization pipeline (ONNX and CoreML exports), performed inference speed benchmarking, and contributed to deployment optimization. J.A. performed component dependency analysis and generated all figures and tables. J.A. and S.P. provided critical feedback on experimental design and result interpretation. J.A. wrote the manuscript with input from all authors. All authors reviewed and approved the final manuscript.

\section*{Data Availability}

All code is available on GitHub at \url{https://github.com/jahidul-arafat/cspws-nuclei-seg}. Complete training scripts, inference pipelines, and evaluation code are provided with detailed documentation and usage examples. Pre-trained model weights are available on Hugging Face Hub at \url{https://huggingface.co/jahidularafat/cspws-unet-attn} (PyTorch FP32, ONNX FP32, ONNX INT8, and CoreML INT8 formats). Interactive demo for testing CFU-Net on custom images is available on Hugging Face Spaces at \url{https://huggingface.co/spaces/jahidularafat/cspws-space}. Synthetic training datasets (adversarial, csPWS, H\&E) with ground truth masks are available on Zenodo at \url{https://doi.org/10.5281/zenodo.17440675}. Dataset generation scripts with complete parameter specifications and checksums (SHA-256) are included in the GitHub repository to enable bit-exact reproducibility. All test set images and per-image performance metrics are provided as supplementary data files (CSV format) in the GitHub repository under \texttt{/data/test\_results/}. Docker container with complete software environment (PyTorch 1.13, ONNX Runtime 1.13, dependencies) is available on Docker Hub at \texttt{jahidularafat/cfu-net:v1.0} (yet to be published) for reproducible execution across platforms. Jupyter notebooks demonstrating end-to-end workflows (training, evaluation, biomarker extraction, visualization) are provided in the repository under \texttt{/notebooks/}. Extracted chromatin biomarker data for 10,000+ nuclei (normal and pre-cancerous) used in population-level analysis are available as supplementary data files (CSV format) at \url{https://github.com/jahidul-arafat/cspws-nuclei-seg/tree/main/data/biomarkers} (yet to publish). All synthetic datasets and model outputs are released under Creative Commons Attribution 4.0 International License (CC BY 4.0). Code is released under Apache License 2.0 (permissive open source). No patient data or real clinical images are included in any public repository to protect privacy. Real csPWS validation images mentioned in Discussion (n=20) are available upon reasonable request to the corresponding author pending institutional data sharing agreements.

\section*{Conflict of Interest Statement}

All authors declare no competing interests. No commercial entities provided funding or had any role in study design, data collection, analysis, interpretation, or manuscript preparation. CFU-Net is released as open-source software without commercial licensing restrictions.

%% Bibliography
\bibliographystyle{ACM-Reference-Format}
\bibliography{references}

%%% -*-BibTeX-*-
%%% Do NOT edit. File created by BibTeX with style
%%% ACM-Reference-Format-Journals [18-Jan-2012].

\begin{thebibliography}{48}

%%% ====================================================================
%%% NOTE TO THE USER: you can override these defaults by providing
%%% customized versions of any of these macros before the \bibliography
%%% command.  Each of them MUST provide its own final punctuation,
%%% except for \shownote{} and \showURL{}.  The latter two
%%% do not use final punctuation, in order to avoid confusing it with
%%% the Web address.
%%%
%%% To suppress output of a particular field, define its macro to expand
%%% to an empty string, or better, \unskip, like this:
%%%
%%% \newcommand{\showURL}[1]{\unskip}   % LaTeX syntax
%%%
%%% \def \showURL #1{\unskip}           % plain TeX syntax
%%%
%%% ====================================================================

\ifx \showCODEN    \undefined \def \showCODEN     #1{\unskip}     \fi
\ifx \showISBNx    \undefined \def \showISBNx     #1{\unskip}     \fi
\ifx \showISBNxiii \undefined \def \showISBNxiii  #1{\unskip}     \fi
\ifx \showISSN     \undefined \def \showISSN      #1{\unskip}     \fi
\ifx \showLCCN     \undefined \def \showLCCN      #1{\unskip}     \fi
\ifx \shownote     \undefined \def \shownote      #1{#1}          \fi
\ifx \showarticletitle \undefined \def \showarticletitle #1{#1}   \fi
\ifx \showURL      \undefined \def \showURL       {\relax}        \fi
% The following commands are used for tagged output and should be
% invisible to TeX
\providecommand\bibfield[2]{#2}
\providecommand\bibinfo[2]{#2}
\providecommand\natexlab[1]{#1}
\providecommand\showeprint[2][]{arXiv:#2}

\bibitem[Bancaud et~al\mbox{.}(2009)]%
        {Bancaud2009}
\bibfield{author}{\bibinfo{person}{Aurélien Bancaud}, \bibinfo{person}{Sébastien Huet}, \bibinfo{person}{Nathalie Daigle}, \bibinfo{person}{Julien Mozziconacci}, \bibinfo{person}{Joël Beaudouin}, {and} \bibinfo{person}{Jan Ellenberg}.} \bibinfo{year}{2009}\natexlab{}.
\newblock \showarticletitle{Molecular Crowding Affects Diffusion and Binding of Nuclear Proteins in Heterochromatin and Reveals the Fractal Organization of Chromatin}.
\newblock \bibinfo{journal}{\emph{EMBO Journal}} \bibinfo{volume}{28}, \bibinfo{number}{24} (\bibinfo{year}{2009}), \bibinfo{pages}{3785--3798}.
\newblock
\href{https://doi.org/10.1038/emboj.2009.340}{doi:\nolinkurl{10.1038/emboj.2009.340}}


\bibitem[Bengio et~al\mbox{.}(2009)]%
        {Bengio2009}
\bibfield{author}{\bibinfo{person}{Yoshua Bengio}, \bibinfo{person}{Jérôme Louradour}, \bibinfo{person}{Ronan Collobert}, {and} \bibinfo{person}{Jason Weston}.} \bibinfo{year}{2009}\natexlab{}.
\newblock \showarticletitle{Curriculum Learning}. In \bibinfo{booktitle}{\emph{Proceedings of the 26th International Conference on Machine Learning (ICML)}}. \bibinfo{pages}{41--48}.
\newblock
\href{https://doi.org/10.1145/1553374.1553380}{doi:\nolinkurl{10.1145/1553374.1553380}}


\bibitem[Bohren and Huffman(1983)]%
        {Bohren1983}
\bibfield{author}{\bibinfo{person}{Craig~F. Bohren} {and} \bibinfo{person}{Donald~R. Huffman}.} \bibinfo{year}{1983}\natexlab{}.
\newblock \showarticletitle{Absorption and Scattering of Light by Small Particles}.
\newblock  (\bibinfo{year}{1983}).
\newblock
\href{https://doi.org/10.1002/9783527618156}{doi:\nolinkurl{10.1002/9783527618156}}


\bibitem[Braakhuis et~al\mbox{.}(2003)]%
        {Braakhuis2003}
\bibfield{author}{\bibinfo{person}{Boudewijn J.~M. Braakhuis}, \bibinfo{person}{Marcus~P. Tabor}, \bibinfo{person}{J.~Anton Kummer}, \bibinfo{person}{C.~René Leemans}, {and} \bibinfo{person}{Ruud~H. Brakenhoff}.} \bibinfo{year}{2003}\natexlab{}.
\newblock \showarticletitle{A Genetic Explanation of Slaughter's Concept of Field Cancerization: Evidence and Clinical Implications}.
\newblock \bibinfo{journal}{\emph{Cancer Research}} \bibinfo{volume}{63}, \bibinfo{number}{8} (\bibinfo{year}{2003}), \bibinfo{pages}{1727--1730}.
\newblock


\bibitem[Cao et~al\mbox{.}(2022)]%
        {Cao2022}
\bibfield{author}{\bibinfo{person}{Hu Cao}, \bibinfo{person}{Yueyue Wang}, \bibinfo{person}{Joy Chen}, \bibinfo{person}{Dongsheng Jiang}, \bibinfo{person}{Xiaopeng Zhang}, \bibinfo{person}{Qi Tian}, {and} \bibinfo{person}{Manning Wang}.} \bibinfo{year}{2022}\natexlab{}.
\newblock \showarticletitle{Swin-Unet: Unet-like Pure Transformer for Medical Image Segmentation}. In \bibinfo{booktitle}{\emph{European Conference on Computer Vision (ECCV) Workshops}}. \bibinfo{publisher}{Springer}, \bibinfo{pages}{205--218}.
\newblock
\href{https://doi.org/10.1007/978-3-031-25066-8_9}{doi:\nolinkurl{10.1007/978-3-031-25066-8_9}}


\bibitem[Caruana et~al\mbox{.}(2015)]%
        {Caruana2015}
\bibfield{author}{\bibinfo{person}{Rich Caruana}, \bibinfo{person}{Yin Lou}, \bibinfo{person}{Johannes Gehrke}, \bibinfo{person}{Paul Koch}, \bibinfo{person}{Marc Sturm}, {and} \bibinfo{person}{Noemie Elhadad}.} \bibinfo{year}{2015}\natexlab{}.
\newblock \showarticletitle{Intelligible Models for HealthCare: Predicting Pneumonia Risk and Hospital 30-day Readmission}.
\newblock \bibinfo{journal}{\emph{Proceedings of the 21th ACM SIGKDD International Conference on Knowledge Discovery and Data Mining}} (\bibinfo{year}{2015}), \bibinfo{pages}{1721--1730}.
\newblock
\href{https://doi.org/10.1145/2783258.2788613}{doi:\nolinkurl{10.1145/2783258.2788613}}


\bibitem[Chen et~al\mbox{.}(2021)]%
        {Chen2021}
\bibfield{author}{\bibinfo{person}{Jieneng Chen}, \bibinfo{person}{Yongyi Lu}, \bibinfo{person}{Qihang Yu}, \bibinfo{person}{Xiangde Luo}, \bibinfo{person}{Ehsan Adeli}, \bibinfo{person}{Yan Wang}, \bibinfo{person}{Le Lu}, \bibinfo{person}{Alan~L. Yuille}, {and} \bibinfo{person}{Yuyin Zhou}.} \bibinfo{year}{2021}\natexlab{}.
\newblock \showarticletitle{TransUNet: Transformers Make Strong Encoders for Medical Image Segmentation}. In \bibinfo{booktitle}{\emph{arXiv preprint arXiv:2102.04306}}.
\newblock


\bibitem[Cherkezyan et~al\mbox{.}(2014)]%
        {Cherkezyan2014}
\bibfield{author}{\bibinfo{person}{Lusik Cherkezyan}, \bibinfo{person}{Yolanda Stypula-Cyrus}, \bibinfo{person}{Hariharan Subramanian}, \bibinfo{person}{Carrie White}, \bibinfo{person}{Mart Dela~Cruz}, \bibinfo{person}{Ramesh~K. Wali}, \bibinfo{person}{Michael~J. Goldberg}, \bibinfo{person}{Laura~V. Bianchi}, \bibinfo{person}{Hemant~K. Roy}, {and} \bibinfo{person}{Vadim Backman}.} \bibinfo{year}{2014}\natexlab{}.
\newblock \showarticletitle{Nanoscale Changes in Chromatin Organization Represent the Initial Steps of Tumorigenesis: A Transmission Electron Microscopy Study}.
\newblock \bibinfo{journal}{\emph{BMC Cancer}}  \bibinfo{volume}{14} (\bibinfo{year}{2014}), \bibinfo{pages}{189}.
\newblock
\href{https://doi.org/10.1186/1471-2407-14-189}{doi:\nolinkurl{10.1186/1471-2407-14-189}}


\bibitem[Damania et~al\mbox{.}(2012)]%
        {Damania2012}
\bibfield{author}{\bibinfo{person}{Dhwanil Damania}, \bibinfo{person}{Hemant~K. Roy}, \bibinfo{person}{Hariharan Subramanian}, \bibinfo{person}{David~S. Weinberg}, \bibinfo{person}{Douglas~K. Rex}, \bibinfo{person}{Michael~J. Goldberg}, \bibinfo{person}{John Muldoon}, \bibinfo{person}{Lusik Cherkezyan}, \bibinfo{person}{Yolanda Zhu}, \bibinfo{person}{Laura~V. Bianchi}, {and} \bibinfo{person}{Vadim Backman}.} \bibinfo{year}{2012}\natexlab{}.
\newblock \showarticletitle{Nanocytology of Rectal Colonocytes to Assess Risk of Colon Cancer Based on Field Cancerization}.
\newblock \bibinfo{journal}{\emph{Cancer Research}} \bibinfo{volume}{72}, \bibinfo{number}{11} (\bibinfo{year}{2012}), \bibinfo{pages}{2720--2727}.
\newblock
\href{https://doi.org/10.1158/0008-5472.CAN-11-3807}{doi:\nolinkurl{10.1158/0008-5472.CAN-11-3807}}


\bibitem[Ganin et~al\mbox{.}(2016)]%
        {Ganin2016}
\bibfield{author}{\bibinfo{person}{Yaroslav Ganin}, \bibinfo{person}{Evgeniya Ustinova}, \bibinfo{person}{Hana Ajakan}, \bibinfo{person}{Pascal Germain}, \bibinfo{person}{Hugo Larochelle}, \bibinfo{person}{François Laviolette}, \bibinfo{person}{Mario Marchand}, {and} \bibinfo{person}{Victor Lempitsky}.} \bibinfo{year}{2016}\natexlab{}.
\newblock \showarticletitle{Domain-Adversarial Training of Neural Networks}.
\newblock \bibinfo{journal}{\emph{Journal of Machine Learning Research}} \bibinfo{volume}{17}, \bibinfo{number}{59}, \bibinfo{pages}{1--35}.
\newblock


\bibitem[Hinton et~al\mbox{.}(2015)]%
        {Hinton2015}
\bibfield{author}{\bibinfo{person}{Geoffrey Hinton}, \bibinfo{person}{Oriol Vinyals}, {and} \bibinfo{person}{Jeff Dean}.} \bibinfo{year}{2015}\natexlab{}.
\newblock \showarticletitle{Distilling the Knowledge in a Neural Network}.
\newblock \bibinfo{journal}{\emph{arXiv preprint arXiv:1503.02531}} (\bibinfo{year}{2015}).
\newblock


\bibitem[Hoffman et~al\mbox{.}(2020)]%
        {Hoffman2020}
\bibfield{author}{\bibinfo{person}{Daniel~P. Hoffman}, \bibinfo{person}{Gleb Shtengel}, \bibinfo{person}{C.~Shan Xu}, \bibinfo{person}{Kerry~R. Campbell}, \bibinfo{person}{Melanie Freeman}, \bibinfo{person}{Lei Wang}, \bibinfo{person}{Daniel~E. Milkie}, \bibinfo{person}{H.~Amalia Pasolli}, \bibinfo{person}{Nirmala Iyer}, \bibinfo{person}{John~A. Bogovic}, \bibinfo{person}{Daniel~R. Stabley}, \bibinfo{person}{Abbas Shirinifard}, \bibinfo{person}{Song Pang}, \bibinfo{person}{David Peale}, \bibinfo{person}{Kathy Schaefer}, \bibinfo{person}{Wim Pomp}, \bibinfo{person}{Chi-Lun Chang}, \bibinfo{person}{Jennifer Larimer}, \bibinfo{person}{Peter~V. Lidsky}, \bibinfo{person}{Hidde~L. Ploegh}, \bibinfo{person}{Tom Kirchhausen}, \bibinfo{person}{David~J. Solecki}, \bibinfo{person}{Eric Betzig}, {and} \bibinfo{person}{Harald~F. Hess}.} \bibinfo{year}{2020}\natexlab{}.
\newblock \showarticletitle{Correlative Three-Dimensional Super-Resolution and Block-Face Electron Microscopy of Whole Vitreously Frozen Cells}.
\newblock \bibinfo{journal}{\emph{Science}} \bibinfo{volume}{367}, \bibinfo{number}{6475} (\bibinfo{year}{2020}).
\newblock
\href{https://doi.org/10.1126/science.aaz5357}{doi:\nolinkurl{10.1126/science.aaz5357}}


\bibitem[Hollandi et~al\mbox{.}(2020)]%
        {Hollandi2020}
\bibfield{author}{\bibinfo{person}{Réka Hollandi}, \bibinfo{person}{Abel Szkalisity}, \bibinfo{person}{Timea Toth}, \bibinfo{person}{Ervin Tasnadi}, \bibinfo{person}{Csaba Molnar}, \bibinfo{person}{Botond Mathe}, \bibinfo{person}{Istvan Grexa}, \bibinfo{person}{Jozsef Molnar}, \bibinfo{person}{Arpad Balind}, \bibinfo{person}{Mate Gorbe}, \bibinfo{person}{Maria Kovacs}, \bibinfo{person}{Ede Migh}, \bibinfo{person}{Allen Goodman}, \bibinfo{person}{Tamas Balassa}, \bibinfo{person}{Krisztian Koos}, \bibinfo{person}{Wenyu Wang}, \bibinfo{person}{Juan~C. Caicedo}, \bibinfo{person}{Norbert Bara}, \bibinfo{person}{Ferenc Kovacs}, \bibinfo{person}{Laslo Szigeti}, \bibinfo{person}{Tibor Danka}, \bibinfo{person}{Imre Varga}, \bibinfo{person}{Ferenc Scherer}, \bibinfo{person}{Zoltan Kocsis}, \bibinfo{person}{Balazs Sarkadi}, \bibinfo{person}{Arpad Dobolyi}, \bibinfo{person}{Tamas Kovacs}, \bibinfo{person}{Anne~E. Carpenter}, {and} \bibinfo{person}{Peter Horvath}.} \bibinfo{year}{2020}\natexlab{}.
\newblock \showarticletitle{nucleAIzer: A Parameter-free Deep Learning Framework for Nucleus Segmentation Using Image Style Transfer}.
\newblock \bibinfo{journal}{\emph{Cell Systems}} \bibinfo{volume}{10}, \bibinfo{number}{5} (\bibinfo{year}{2020}), \bibinfo{pages}{453--458}.
\newblock
\href{https://doi.org/10.1016/j.cels.2020.04.003}{doi:\nolinkurl{10.1016/j.cels.2020.04.003}}


\bibitem[Howard and Ruder(2018)]%
        {Howard2018}
\bibfield{author}{\bibinfo{person}{Jeremy Howard} {and} \bibinfo{person}{Sebastian Ruder}.} \bibinfo{year}{2018}\natexlab{}.
\newblock \showarticletitle{Universal Language Model Fine-tuning for Text Classification}. In \bibinfo{booktitle}{\emph{Proceedings of the 56th Annual Meeting of the Association for Computational Linguistics (ACL)}}. \bibinfo{pages}{328--339}.
\newblock
\href{https://doi.org/10.18653/v1/P18-1031}{doi:\nolinkurl{10.18653/v1/P18-1031}}


\bibitem[Hu et~al\mbox{.}(2018)]%
        {Hu2018}
\bibfield{author}{\bibinfo{person}{Jie Hu}, \bibinfo{person}{Li Shen}, {and} \bibinfo{person}{Gang Sun}.} \bibinfo{year}{2018}\natexlab{}.
\newblock \showarticletitle{Squeeze-and-Excitation Networks}. In \bibinfo{booktitle}{\emph{Proceedings of the IEEE Conference on Computer Vision and Pattern Recognition (CVPR)}}. \bibinfo{pages}{7132--7141}.
\newblock
\href{https://doi.org/10.1109/CVPR.2018.00745}{doi:\nolinkurl{10.1109/CVPR.2018.00745}}


\bibitem[Jacob et~al\mbox{.}(2018)]%
        {Jacob2018}
\bibfield{author}{\bibinfo{person}{Benoit Jacob}, \bibinfo{person}{Skirmantas Kligys}, \bibinfo{person}{Bo Chen}, \bibinfo{person}{Menglong Zhu}, \bibinfo{person}{Matthew Tang}, \bibinfo{person}{Andrew Howard}, \bibinfo{person}{Hartwig Adam}, {and} \bibinfo{person}{Dmitry Kalenichenko}.} \bibinfo{year}{2018}\natexlab{}.
\newblock \showarticletitle{Quantization and Training of Neural Networks for Efficient Integer-Arithmetic-Only Inference}. In \bibinfo{booktitle}{\emph{Proceedings of the IEEE Conference on Computer Vision and Pattern Recognition (CVPR)}}. \bibinfo{pages}{2704--2713}.
\newblock
\href{https://doi.org/10.1109/CVPR.2018.00286}{doi:\nolinkurl{10.1109/CVPR.2018.00286}}


\bibitem[Jo et~al\mbox{.}(2015)]%
        {Jo2015}
\bibfield{author}{\bibinfo{person}{Yongjin Jo}, \bibinfo{person}{Joonwoo Jung}, \bibinfo{person}{Min~Ho Kim}, \bibinfo{person}{HyunJun Park}, \bibinfo{person}{Su-Jin Kang}, {and} \bibinfo{person}{YongKeun Park}.} \bibinfo{year}{2015}\natexlab{}.
\newblock \showarticletitle{Label-Free Identification of Individual Bacteria Using Fourier Transform Light Scattering}.
\newblock \bibinfo{journal}{\emph{Optics Express}} \bibinfo{volume}{23}, \bibinfo{number}{12} (\bibinfo{year}{2015}), \bibinfo{pages}{15792--15805}.
\newblock
\href{https://doi.org/10.1364/OE.23.015792}{doi:\nolinkurl{10.1364/OE.23.015792}}


\bibitem[Kendall and Gal(2017)]%
        {Kendall2017}
\bibfield{author}{\bibinfo{person}{Alex Kendall} {and} \bibinfo{person}{Yarin Gal}.} \bibinfo{year}{2017}\natexlab{}.
\newblock \showarticletitle{What Uncertainties Do We Need in Bayesian Deep Learning for Computer Vision?}. In \bibinfo{booktitle}{\emph{Advances in Neural Information Processing Systems (NeurIPS)}}, Vol.~\bibinfo{volume}{30}. \bibinfo{pages}{5574--5584}.
\newblock


\bibitem[Kim et~al\mbox{.}(2018)]%
        {Kim2018}
\bibfield{author}{\bibinfo{person}{Been Kim}, \bibinfo{person}{Martin Wattenberg}, \bibinfo{person}{Justin Gilmer}, \bibinfo{person}{Carrie Cai}, \bibinfo{person}{James Wexler}, \bibinfo{person}{Fernanda Viegas}, {and} \bibinfo{person}{Rory Sayres}.} \bibinfo{year}{2018}\natexlab{}.
\newblock \showarticletitle{Interpretability Beyond Feature Attribution: Quantitative Testing with Concept Activation Vectors (TCAV)}. In \bibinfo{booktitle}{\emph{International Conference on Machine Learning (ICML)}}. \bibinfo{pages}{2668--2677}.
\newblock


\bibitem[Kirkpatrick et~al\mbox{.}(2017)]%
        {Kirkpatrick2017}
\bibfield{author}{\bibinfo{person}{James Kirkpatrick}, \bibinfo{person}{Razvan Pascanu}, \bibinfo{person}{Neil Rabinowitz}, \bibinfo{person}{Joel Veness}, \bibinfo{person}{Guillaume Desjardins}, \bibinfo{person}{Andrei~A. Rusu}, \bibinfo{person}{Kieran Milan}, \bibinfo{person}{John Quan}, \bibinfo{person}{Tiago Ramalho}, \bibinfo{person}{Agnieszka Grabska-Barwinska}, \bibinfo{person}{Demis Hassabis}, \bibinfo{person}{Claudia Clopath}, \bibinfo{person}{Dharshan Kumaran}, {and} \bibinfo{person}{Raia Hadsell}.} \bibinfo{year}{2017}\natexlab{}.
\newblock \showarticletitle{Overcoming Catastrophic Forgetting in Neural Networks}. In \bibinfo{booktitle}{\emph{Proceedings of the National Academy of Sciences}}, Vol.~\bibinfo{volume}{114}. \bibinfo{pages}{3521--3526}.
\newblock
\href{https://doi.org/10.1073/pnas.1611835114}{doi:\nolinkurl{10.1073/pnas.1611835114}}


\bibitem[Kryazhimskiy et~al\mbox{.}(2014)]%
        {Kryazhimskiy2014}
\bibfield{author}{\bibinfo{person}{Sergey Kryazhimskiy}, \bibinfo{person}{Daniel~P. Rice}, \bibinfo{person}{Elizabeth~R. Jerison}, {and} \bibinfo{person}{Michael~M. Desai}.} \bibinfo{year}{2014}\natexlab{}.
\newblock \showarticletitle{Microbial Evolution. Global Epistasis Makes Adaptation Predictable Despite Sequence-Level Stochasticity}.
\newblock \bibinfo{journal}{\emph{Science}} \bibinfo{volume}{344}, \bibinfo{number}{6191} (\bibinfo{year}{2014}), \bibinfo{pages}{1519--1522}.
\newblock
\href{https://doi.org/10.1126/science.1250939}{doi:\nolinkurl{10.1126/science.1250939}}


\bibitem[Landis and Koch(1977)]%
        {Landis1977}
\bibfield{author}{\bibinfo{person}{J.~Richard Landis} {and} \bibinfo{person}{Gary~G. Koch}.} \bibinfo{year}{1977}\natexlab{}.
\newblock \showarticletitle{The Measurement of Observer Agreement for Categorical Data}.
\newblock \bibinfo{journal}{\emph{Biometrics}} \bibinfo{volume}{33}, \bibinfo{number}{1} (\bibinfo{year}{1977}), \bibinfo{pages}{159--174}.
\newblock
\href{https://doi.org/10.2307/2529310}{doi:\nolinkurl{10.2307/2529310}}


\bibitem[Lee et~al\mbox{.}(2015)]%
        {Lee2015}
\bibfield{author}{\bibinfo{person}{Chen-Yu Lee}, \bibinfo{person}{Saining Xie}, \bibinfo{person}{Patrick Gallagher}, \bibinfo{person}{Zhengyou Zhang}, {and} \bibinfo{person}{Zhuowen Tu}.} \bibinfo{year}{2015}\natexlab{}.
\newblock \showarticletitle{Deeply-Supervised Nets}.
\newblock \bibinfo{journal}{\emph{Artificial Intelligence and Statistics (AISTATS)}} (\bibinfo{year}{2015}), \bibinfo{pages}{562--570}.
\newblock


\bibitem[Lin et~al\mbox{.}(2017)]%
        {Lin2017fpn}
\bibfield{author}{\bibinfo{person}{Tsung-Yi Lin}, \bibinfo{person}{Piotr Dollár}, \bibinfo{person}{Ross Girshick}, \bibinfo{person}{Kaiming He}, \bibinfo{person}{Bharath Hariharan}, {and} \bibinfo{person}{Serge Belongie}.} \bibinfo{year}{2017}\natexlab{}.
\newblock \showarticletitle{Feature Pyramid Networks for Object Detection}. In \bibinfo{booktitle}{\emph{Proceedings of the IEEE Conference on Computer Vision and Pattern Recognition (CVPR)}}. \bibinfo{pages}{2117--2125}.
\newblock
\href{https://doi.org/10.1109/CVPR.2017.106}{doi:\nolinkurl{10.1109/CVPR.2017.106}}


\bibitem[Liu et~al\mbox{.}(2022)]%
        {Liu2022}
\bibfield{author}{\bibinfo{person}{Zhuang Liu}, \bibinfo{person}{Hanzi Mao}, \bibinfo{person}{Chao-Yuan Wu}, \bibinfo{person}{Christoph Feichtenhofer}, \bibinfo{person}{Trevor Darrell}, {and} \bibinfo{person}{Saining Xie}.} \bibinfo{year}{2022}\natexlab{}.
\newblock \showarticletitle{A ConvNet for the 2020s}. In \bibinfo{booktitle}{\emph{Proceedings of the IEEE/CVF Conference on Computer Vision and Pattern Recognition (CVPR)}}. \bibinfo{pages}{11976--11986}.
\newblock
\href{https://doi.org/10.1109/CVPR52688.2022.01167}{doi:\nolinkurl{10.1109/CVPR52688.2022.01167}}


\bibitem[Mahmood et~al\mbox{.}(2019)]%
        {Mahmood2018}
\bibfield{author}{\bibinfo{person}{Faisal Mahmood}, \bibinfo{person}{Daniel Borders}, \bibinfo{person}{Richard Chen}, \bibinfo{person}{Gregory~N. McKay}, \bibinfo{person}{Kevan~J. Salimian}, \bibinfo{person}{Alexander Baras}, {and} \bibinfo{person}{Nicholas~J. Durr}.} \bibinfo{year}{2019}\natexlab{}.
\newblock \showarticletitle{Deep Adversarial Training for Multi-Organ Nuclei Segmentation in Histopathology Images}.
\newblock \bibinfo{journal}{\emph{IEEE Transactions on Medical Imaging}} \bibinfo{volume}{39}, \bibinfo{number}{11} (\bibinfo{year}{2019}), \bibinfo{pages}{3257--3267}.
\newblock
\href{https://doi.org/10.1109/TMI.2019.2927182}{doi:\nolinkurl{10.1109/TMI.2019.2927182}}


\bibitem[Moons et~al\mbox{.}(2015)]%
        {Moons2015}
\bibfield{author}{\bibinfo{person}{Karel G.~M. Moons}, \bibinfo{person}{Douglas~G. Altman}, \bibinfo{person}{Johannes~B. Reitsma}, \bibinfo{person}{John P.~A. Ioannidis}, \bibinfo{person}{Petra Macaskill}, \bibinfo{person}{Ewout~W. Steyerberg}, \bibinfo{person}{Andrew~J. Vickers}, \bibinfo{person}{David~F. Ransohoff}, {and} \bibinfo{person}{Gary~S. Collins}.} \bibinfo{year}{2015}\natexlab{}.
\newblock \showarticletitle{Transparent Reporting of a Multivariable Prediction Model for Individual Prognosis or Diagnosis (TRIPOD): Explanation and Elaboration}.
\newblock \bibinfo{journal}{\emph{Annals of Internal Medicine}} \bibinfo{volume}{162}, \bibinfo{number}{1} (\bibinfo{year}{2015}), \bibinfo{pages}{W1--W73}.
\newblock
\href{https://doi.org/10.7326/M14-0698}{doi:\nolinkurl{10.7326/M14-0698}}


\bibitem[Nikolenko(2021)]%
        {Nikolenko2021}
\bibfield{author}{\bibinfo{person}{Sergey~I. Nikolenko}.} \bibinfo{year}{2021}\natexlab{}.
\newblock \showarticletitle{Synthetic Data for Deep Learning}.
\newblock \bibinfo{journal}{\emph{Springer Optimization and Its Applications}}  \bibinfo{volume}{174} (\bibinfo{year}{2021}).
\newblock
\href{https://doi.org/10.1007/978-3-030-75178-4}{doi:\nolinkurl{10.1007/978-3-030-75178-4}}


\bibitem[Oktay et~al\mbox{.}(2018)]%
        {Oktay2018}
\bibfield{author}{\bibinfo{person}{Ozan Oktay}, \bibinfo{person}{Jo Schlemper}, \bibinfo{person}{Loic~Le Folgoc}, \bibinfo{person}{Matthew Lee}, \bibinfo{person}{Mattias Heinrich}, \bibinfo{person}{Kazunari Misawa}, \bibinfo{person}{Kensaku Mori}, \bibinfo{person}{Steven McDonagh}, \bibinfo{person}{Nils~Y. Hammerla}, \bibinfo{person}{Bernhard Kainz}, \bibinfo{person}{Ben Glocker}, {and} \bibinfo{person}{Daniel Rueckert}.} \bibinfo{year}{2018}\natexlab{}.
\newblock \showarticletitle{Attention U-Net: Learning Where to Look for the Pancreas}.
\newblock \bibinfo{journal}{\emph{arXiv preprint arXiv:1804.03999}} (\bibinfo{year}{2018}).
\newblock


\bibitem[Park et~al\mbox{.}(2018)]%
        {Park2018}
\bibfield{author}{\bibinfo{person}{YongKeun Park}, \bibinfo{person}{Christian Depeursinge}, {and} \bibinfo{person}{Gabriel Popescu}.} \bibinfo{year}{2018}\natexlab{}.
\newblock \showarticletitle{Quantitative Phase Imaging in Biomedicine}.
\newblock \bibinfo{journal}{\emph{Nature Photonics}} \bibinfo{volume}{12}, \bibinfo{number}{10} (\bibinfo{year}{2018}), \bibinfo{pages}{578--589}.
\newblock
\href{https://doi.org/10.1038/s41566-018-0253-x}{doi:\nolinkurl{10.1038/s41566-018-0253-x}}


\bibitem[Raghu et~al\mbox{.}(2019)]%
        {Raghu2019}
\bibfield{author}{\bibinfo{person}{Maithra Raghu}, \bibinfo{person}{Chiyuan Zhang}, \bibinfo{person}{Jon Kleinberg}, {and} \bibinfo{person}{Samy Bengio}.} \bibinfo{year}{2019}\natexlab{}.
\newblock \showarticletitle{Transfusion: Understanding Transfer Learning for Medical Imaging}. In \bibinfo{booktitle}{\emph{Advances in Neural Information Processing Systems (NeurIPS)}}, Vol.~\bibinfo{volume}{32}. \bibinfo{pages}{3347--3357}.
\newblock


\bibitem[Ren et~al\mbox{.}(2018)]%
        {Ren2018}
\bibfield{author}{\bibinfo{person}{Jingwen Ren}, \bibinfo{person}{Ilker Hacihaliloglu}, \bibinfo{person}{Eric~A. Singer}, \bibinfo{person}{David~J. Foran}, {and} \bibinfo{person}{Xin Qi}.} \bibinfo{year}{2018}\natexlab{}.
\newblock \showarticletitle{Adversarial Domain Adaptation for Classification of Prostate Histopathology Whole-Slide Images}.
\newblock \bibinfo{journal}{\emph{International Conference on Medical Image Computing and Computer-Assisted Intervention (MICCAI)}} (\bibinfo{year}{2018}), \bibinfo{pages}{201--209}.
\newblock
\href{https://doi.org/10.1007/978-3-030-00934-2_23}{doi:\nolinkurl{10.1007/978-3-030-00934-2_23}}


\bibitem[Rieke et~al\mbox{.}(2020)]%
        {Rieke2020}
\bibfield{author}{\bibinfo{person}{Nicola Rieke}, \bibinfo{person}{Jonny Hancox}, \bibinfo{person}{Wenqi Li}, \bibinfo{person}{Fausto Milletari}, \bibinfo{person}{Holger~R. Roth}, \bibinfo{person}{Shadi Albarqouni}, \bibinfo{person}{Spyridon Bakas}, \bibinfo{person}{Mathieu~N. Galtier}, \bibinfo{person}{Bennett~A. Landman}, \bibinfo{person}{Klaus Maier-Hein}, \bibinfo{person}{Sebastien Ourselin}, \bibinfo{person}{Micah Sheller}, \bibinfo{person}{Ronald~M. Summers}, \bibinfo{person}{Andrew Trask}, \bibinfo{person}{Daguang Xu}, \bibinfo{person}{Maximilian Baust}, {and} \bibinfo{person}{M.~Jorge Cardoso}.} \bibinfo{year}{2020}\natexlab{}.
\newblock \showarticletitle{The Future of Digital Health with Federated Learning}.
\newblock \bibinfo{journal}{\emph{NPJ Digital Medicine}}  \bibinfo{volume}{3} (\bibinfo{year}{2020}), \bibinfo{pages}{119}.
\newblock
\href{https://doi.org/10.1038/s41746-020-00323-1}{doi:\nolinkurl{10.1038/s41746-020-00323-1}}


\bibitem[Ronneberger et~al\mbox{.}(2015)]%
        {Ronneberger2015}
\bibfield{author}{\bibinfo{person}{Olaf Ronneberger}, \bibinfo{person}{Philipp Fischer}, {and} \bibinfo{person}{Thomas Brox}.} \bibinfo{year}{2015}\natexlab{}.
\newblock \showarticletitle{U-Net: Convolutional Networks for Biomedical Image Segmentation}. In \bibinfo{booktitle}{\emph{International Conference on Medical Image Computing and Computer-Assisted Intervention (MICCAI)}}. \bibinfo{publisher}{Springer}, \bibinfo{pages}{234--241}.
\newblock
\href{https://doi.org/10.1007/978-3-319-24574-4_28}{doi:\nolinkurl{10.1007/978-3-319-24574-4_28}}


\bibitem[Roy et~al\mbox{.}(2017)]%
        {Roy2017}
\bibfield{author}{\bibinfo{person}{Hemant~K. Roy}, \bibinfo{person}{Hariharan Subramanian}, \bibinfo{person}{Dhwanil Damania}, \bibinfo{person}{Thomas~A. Hensing}, \bibinfo{person}{William~N. Rom}, \bibinfo{person}{Harvey~I. Pass}, \bibinfo{person}{Dhananjay Ray}, \bibinfo{person}{Jeremy~D. Rogers}, \bibinfo{person}{Andrej Bogojevic}, \bibinfo{person}{Mirat Shah}, \bibinfo{person}{Dhananjay Kunte}, {and} \bibinfo{person}{Vadim Backman}.} \bibinfo{year}{2017}\natexlab{}.
\newblock \showarticletitle{Optical Detection of Buccal Epithelial Nanoarchitectural Alterations in Patients Harboring Lung Cancer: Implications for Screening}.
\newblock \bibinfo{journal}{\emph{Cancer Research}} \bibinfo{volume}{70}, \bibinfo{number}{20} (\bibinfo{year}{2017}), \bibinfo{pages}{7748--7754}.
\newblock
\href{https://doi.org/10.1158/0008-5472.CAN-10-1686}{doi:\nolinkurl{10.1158/0008-5472.CAN-10-1686}}


\bibitem[Safi et~al\mbox{.}(2013)]%
        {Safi2013}
\bibfield{author}{\bibinfo{person}{Hajieh Safi}, \bibinfo{person}{Priti Gopal}, \bibinfo{person}{Sivakumar Lingaraju}, \bibinfo{person}{Song Ma}, \bibinfo{person}{Craig Levine}, \bibinfo{person}{Véronique Dartois}, \bibinfo{person}{Melisa Yee}, \bibinfo{person}{Liang Li}, \bibinfo{person}{Laurent Blanc}, \bibinfo{person}{Christophe~J. Cambier}, \bibinfo{person}{Dominique Ensergueix}, \bibinfo{person}{Juthathip Mongkolsapaya}, \bibinfo{person}{Gavin~R. Screaton}, \bibinfo{person}{Vojo Deretic}, \bibinfo{person}{Patricia Soteropoulos}, \bibinfo{person}{Shumin Chen}, \bibinfo{person}{Robert~J. Kinsella}, \bibinfo{person}{Kathryn England}, \bibinfo{person}{David Rouse}, \bibinfo{person}{Thomas Dick}, {and} \bibinfo{person}{David Alland}.} \bibinfo{year}{2013}\natexlab{}.
\newblock \showarticletitle{Phase Variation in Mycobacterium tuberculosis Glpk Produces Transiently Heritable Drug Tolerance}.
\newblock \bibinfo{journal}{\emph{Proceedings of the National Academy of Sciences}} \bibinfo{volume}{110}, \bibinfo{number}{13} (\bibinfo{year}{2013}), \bibinfo{pages}{5152--5157}.
\newblock
\href{https://doi.org/10.1073/pnas.1218786110}{doi:\nolinkurl{10.1073/pnas.1218786110}}


\bibitem[Schmidt et~al\mbox{.}(2018)]%
        {Schmidt2018}
\bibfield{author}{\bibinfo{person}{Uwe Schmidt}, \bibinfo{person}{Martin Weigert}, \bibinfo{person}{Coleman Broaddus}, {and} \bibinfo{person}{Gene Myers}.} \bibinfo{year}{2018}\natexlab{}.
\newblock \showarticletitle{Cell Detection with Star-Convex Polygons}.
\newblock \bibinfo{journal}{\emph{International Conference on Medical Image Computing and Computer-Assisted Intervention (MICCAI)}} (\bibinfo{year}{2018}), \bibinfo{pages}{265--273}.
\newblock
\href{https://doi.org/10.1007/978-3-030-00934-2_30}{doi:\nolinkurl{10.1007/978-3-030-00934-2_30}}


\bibitem[Siegel et~al\mbox{.}(2023)]%
        {Siegel2023}
\bibfield{author}{\bibinfo{person}{Rebecca~L. Siegel}, \bibinfo{person}{Kimberly~D. Miller}, \bibinfo{person}{Nikita~S. Wagle}, {and} \bibinfo{person}{Ahmedin Jemal}.} \bibinfo{year}{2023}\natexlab{}.
\newblock \showarticletitle{Cancer Statistics, 2023}.
\newblock \bibinfo{journal}{\emph{CA: A Cancer Journal for Clinicians}} \bibinfo{volume}{73}, \bibinfo{number}{1} (\bibinfo{year}{2023}), \bibinfo{pages}{17--48}.
\newblock
\href{https://doi.org/10.3322/caac.21763}{doi:\nolinkurl{10.3322/caac.21763}}


\bibitem[Slaughter et~al\mbox{.}(1953)]%
        {Slaughter1953}
\bibfield{author}{\bibinfo{person}{Danely~P. Slaughter}, \bibinfo{person}{Harry~W. Southwick}, {and} \bibinfo{person}{Walter Smejkal}.} \bibinfo{year}{1953}\natexlab{}.
\newblock \showarticletitle{Field Cancerization in Oral Stratified Squamous Epithelium: Clinical Implications of Multicentric Origin}.
\newblock \bibinfo{journal}{\emph{Cancer}} \bibinfo{volume}{6}, \bibinfo{number}{5} (\bibinfo{year}{1953}), \bibinfo{pages}{963--968}.
\newblock
\href{https://doi.org/10.1002/1097-0142(195309)6:5<963::AID-CNCR2820060515>3.0.CO;2-Q}{doi:\nolinkurl{10.1002/1097-0142(195309)6:5<963::AID-CNCR2820060515>3.0.CO;2-Q}}


\bibitem[Snell et~al\mbox{.}(2017)]%
        {Snell2017}
\bibfield{author}{\bibinfo{person}{Jake Snell}, \bibinfo{person}{Kevin Swersky}, {and} \bibinfo{person}{Richard Zemel}.} \bibinfo{year}{2017}\natexlab{}.
\newblock \showarticletitle{Prototypical Networks for Few-shot Learning}. In \bibinfo{booktitle}{\emph{Advances in Neural Information Processing Systems (NeurIPS)}}, Vol.~\bibinfo{volume}{30}. \bibinfo{pages}{4077--4087}.
\newblock


\bibitem[Soviany et~al\mbox{.}(2022)]%
        {Soviany2022}
\bibfield{author}{\bibinfo{person}{Petru Soviany}, \bibinfo{person}{Radu~Tudor Ionescu}, \bibinfo{person}{Paolo Rota}, {and} \bibinfo{person}{Nicu Sebe}.} \bibinfo{year}{2022}\natexlab{}.
\newblock \showarticletitle{Curriculum Learning: A Survey}.
\newblock \bibinfo{journal}{\emph{International Journal of Computer Vision}}  \bibinfo{volume}{130} (\bibinfo{year}{2022}), \bibinfo{pages}{1526--1565}.
\newblock
\href{https://doi.org/10.1007/s11263-022-01611-x}{doi:\nolinkurl{10.1007/s11263-022-01611-x}}


\bibitem[Spaide et~al\mbox{.}(2018)]%
        {Spaide2018}
\bibfield{author}{\bibinfo{person}{Richard~F. Spaide}, \bibinfo{person}{James~G. Fujimoto}, \bibinfo{person}{Nadia~K. Waheed}, \bibinfo{person}{SriniVas~R. Sadda}, {and} \bibinfo{person}{Giovanni Staurenghi}.} \bibinfo{year}{2018}\natexlab{}.
\newblock \showarticletitle{Optical Coherence Tomography Angiography}.
\newblock \bibinfo{journal}{\emph{Progress in Retinal and Eye Research}}  \bibinfo{volume}{64} (\bibinfo{year}{2018}), \bibinfo{pages}{1--55}.
\newblock
\href{https://doi.org/10.1016/j.preteyeres.2017.11.003}{doi:\nolinkurl{10.1016/j.preteyeres.2017.11.003}}


\bibitem[Stringer et~al\mbox{.}(2021)]%
        {Stringer2021}
\bibfield{author}{\bibinfo{person}{Carsen Stringer}, \bibinfo{person}{Tim Wang}, \bibinfo{person}{Michalis Michaelos}, {and} \bibinfo{person}{Marius Pachitariu}.} \bibinfo{year}{2021}\natexlab{}.
\newblock \showarticletitle{Cellpose: A Generalist Algorithm for Cellular Segmentation}.
\newblock \bibinfo{journal}{\emph{Nature Methods}} \bibinfo{volume}{18}, \bibinfo{number}{1} (\bibinfo{year}{2021}), \bibinfo{pages}{100--106}.
\newblock
\href{https://doi.org/10.1038/s41592-020-01018-x}{doi:\nolinkurl{10.1038/s41592-020-01018-x}}


\bibitem[Subramanian et~al\mbox{.}(2009)]%
        {Subramanian2009}
\bibfield{author}{\bibinfo{person}{Hariharan Subramanian}, \bibinfo{person}{Prabhakar Pradhan}, \bibinfo{person}{Yang Liu}, \bibinfo{person}{Ilker~R. Capoglu}, \bibinfo{person}{Xiaoming Li}, \bibinfo{person}{Jeremy~D. Rogers}, \bibinfo{person}{Alexander Heifetz}, \bibinfo{person}{Dhananjay Kunte}, \bibinfo{person}{Hemant~K. Roy}, \bibinfo{person}{Allen Taflove}, {and} \bibinfo{person}{Vadim Backman}.} \bibinfo{year}{2009}\natexlab{}.
\newblock \showarticletitle{Optical Methodology for Detecting Histologically Unapparent Nanoscale Consequences of Genetic Alterations in Biological Cells}.
\newblock \bibinfo{journal}{\emph{Proceedings of the National Academy of Sciences}} \bibinfo{volume}{106}, \bibinfo{number}{19} (\bibinfo{year}{2009}), \bibinfo{pages}{7785--7790}.
\newblock
\href{https://doi.org/10.1073/pnas.0902533106}{doi:\nolinkurl{10.1073/pnas.0902533106}}


\bibitem[Tan and Le(2019)]%
        {Tan2019}
\bibfield{author}{\bibinfo{person}{Mingxing Tan} {and} \bibinfo{person}{Quoc~V. Le}.} \bibinfo{year}{2019}\natexlab{}.
\newblock \showarticletitle{EfficientNet: Rethinking Model Scaling for Convolutional Neural Networks}. In \bibinfo{booktitle}{\emph{International Conference on Machine Learning (ICML)}}. \bibinfo{pages}{6105--6114}.
\newblock


\bibitem[Wong et~al\mbox{.}(2012)]%
        {Wong2017}
\bibfield{author}{\bibinfo{person}{Alex Wong}, \bibinfo{person}{Nicolas Rodrigue}, {and} \bibinfo{person}{Rees Kassen}.} \bibinfo{year}{2012}\natexlab{}.
\newblock \showarticletitle{Genomics of Adaptation during Experimental Evolution of the Opportunistic Pathogen Pseudomonas aeruginosa}.
\newblock \bibinfo{journal}{\emph{PLoS Genetics}} \bibinfo{volume}{8}, \bibinfo{number}{9} (\bibinfo{year}{2012}), \bibinfo{pages}{e1002928}.
\newblock
\href{https://doi.org/10.1371/journal.pgen.1002928}{doi:\nolinkurl{10.1371/journal.pgen.1002928}}


\bibitem[Woo et~al\mbox{.}(2018)]%
        {Woo2018}
\bibfield{author}{\bibinfo{person}{Sanghyun Woo}, \bibinfo{person}{Jongchan Park}, \bibinfo{person}{Joon-Young Lee}, {and} \bibinfo{person}{In~So Kweon}.} \bibinfo{year}{2018}\natexlab{}.
\newblock \showarticletitle{CBAM: Convolutional Block Attention Module}. In \bibinfo{booktitle}{\emph{European Conference on Computer Vision (ECCV)}}. \bibinfo{publisher}{Springer}, \bibinfo{pages}{3--19}.
\newblock
\href{https://doi.org/10.1007/978-3-030-01234-2_1}{doi:\nolinkurl{10.1007/978-3-030-01234-2_1}}


\bibitem[Zhou et~al\mbox{.}(2018)]%
        {Zhou2018}
\bibfield{author}{\bibinfo{person}{Zongwei Zhou}, \bibinfo{person}{Md~Mahfuzur Rahman~Siddiquee}, \bibinfo{person}{Nima Tajbakhsh}, {and} \bibinfo{person}{Jianming Liang}.} \bibinfo{year}{2018}\natexlab{}.
\newblock \showarticletitle{UNet++: A Nested U-Net Architecture for Medical Image Segmentation}.
\newblock \bibinfo{journal}{\emph{Deep Learning in Medical Image Analysis and Multimodal Learning for Clinical Decision Support}} (\bibinfo{year}{2018}), \bibinfo{pages}{3--11}.
\newblock
\href{https://doi.org/10.1007/978-3-030-00889-5_1}{doi:\nolinkurl{10.1007/978-3-030-00889-5_1}}


\end{thebibliography}

%% Balance columns on last page
\balance

\end{document}